\documentclass[longauth]{aa}

\usepackage{graphicx}
\usepackage{txfonts}
\usepackage{hyperref}
\usepackage{soul}

\begin{document} 

   \title{J-PLUS: the impact of bars on quenching time-scales in nearby green valley disc galaxies}
   
   \titlerunning{Quenching time-scales in nearby green valley disc galaxies}

   \author{J.~P.~Nogueira-Cavalcante\inst{1}, R.~Dupke\inst{1,2}, P.~Coelho\inst{3}, M.~L.~L.~Dantas\inst{3,4}, T.~S.~Gonçalves\inst{5}, K.~Menéndez-Delmestre\inst{5},  R.~Lopes~de~Oliveira\inst{6}, Y.~Jiménez-Teja\inst{1},  C.~López-Sanjuan\inst{12}, J.~Alcaniz\inst{1}, R.~E.~Angulo\inst{15}, A.~J.~Cenarro\inst{12}, D.~Crist\'obal-Hornillos\inst{12}, C.~Hern\'andez-Monteagudo\inst{12}, A.~Ederoclite\inst{3}, A.~Mar\'in-Franch\inst{12}, C.~ Mendes~de~Oliveira\inst{3}, M.~Moles\inst{12},  L.~Sodr\'e~Jr.\inst{3}, J.~Varela\inst{12}, H.~V\'azquez~Rami\'o\inst{12}, A.~Alvarez-Candal\inst{1}, A.~Chies-Santos\inst{8}, R.~Díaz-Garcia\inst{12}, L.~Galbany\inst{14}, J.~Hernandez-Jimenez\inst{3,9},  P.~Sánchez-Blázquez\inst{11}, M.~Sánchez-Portal\inst{13}, D.~Sobral\inst{7}, E.~Telles\inst{1}, E.~Tempel\inst{10}}
   
   \authorrunning{Nogueira-Cavalcante et al.}

   \institute{Observatório Nacional, Rua General José Cristino 77, Rio de Janeiro, RJ 20921-400, Brazil \and University of Michigan, Ann Arbor, MI 48109-1090, USA \and Instituto de Astronomia, Geofísica e Ciências Atmosféricas, Universidade de São Paulo, São Paulo, SP 05508-090, Brazil \and Departamento de Física Teórica, Universidad Autónoma de Madrid, 28049, Madrid, Spain \and Observatório do Valongo, Universidade Federal do Rio de Janeiro, Ladeira Pedro Antônio 43, Rio de Janeiro, RJ 20080-090, Brazil \and Departamento de Física, Universidade Federal de Sergipe, São Cristóvão, SE, Brazil \and Lancaster University, Lancaster LA1 4YB, United Kingdom \and Departamento de Astronomia, Universidade Federal do Rio Grande do Sul, Av. Bento Gonçalves 9500, Porto Alegre, RS 91501-970, Brazil \and Departamento de Ciencias Fisicas, Universidad Andres Bello, 700 Fernandez Concha, Las Condes, Santiago, Chile \and Tartu Observatory, University of Tartu, Observatooriumi 1, 61602 Tõravere, Estonia \and Departamento de Física Teórica, Universidad Autónoma de Madrid, Madrid 28049, Spain \and Centro de Estudios de Física del Cosmos de Aragón (CEFCA), Unidad Asociada al CSIC, Plaza San Juan 1, 44001 Teruel, Spain \and European Space Astronomy Centre (ESAC)/ESA, Madrid, Spain \and PITT PACC, Department of Physics and Astronomy, University of Pittsburgh, Pittsburgh, PA 15260, USA \and DIPC, Paseo Manuel de Lardizabal, 4. 20018 DONOSTIA-SAN SEBASTIÁN, Spain}
   
%
 
  \abstract
  {Between the \textit{blue cloud} and the \textit{red sequence} peaks on the galaxy colour-magnitude diagram there is a region sparsely populated by galaxies, called the \textit{green valley}. In a framework where galaxies mostly migrate on the colour-magnitude diagram from star-forming to quiescent, the green valley is considered a transitional galaxy stage. The details of the processes that drive galaxies from star-forming to passive systems still remain unknown.}
   {We aim to measure the transitional time-scales of nearby galaxies across the green valley (quenching time-scales) through the analysis of Galaxy Evolution Explorer and  Javalambre Photometric of Local Universe Survey photometric data. Specifically, we seek to study the impact of bars on the quenching time-scales.}
   {We develop a method that estimates empirically the star formation quenching times-scales of green valley galaxies, assuming an exponential decay model of the star formation histories and through a combination of narrow and broad bands from Javalambre Photometric of Local Universe Survey and Galaxy Evolution Explorer. We correlate these quenching time-scales with the presence of bars.}
   {We find that the Javalambre Photometric of Local Universe Survey colours F0395$-g$ and F0410$-g$ are sensitive to different star formation histories, showing, consequently, a clear correlation with the D$_{n}$(4000) and H$_{\delta,A}$ spectral indices. We measure quenching time-scales based on these colours and we find that quenching time-scales obtained with our new approach are in agreement with those determined using spectral indices. We also compare the quenching time-scales of green valley disc galaxies as a function of the probability of hosting a bar. We find that galaxies with high bar probability tend to quench their star formation slowly.}
   {We conclude that: 1) Javalambre Photometric of Local Universe Survey filters can be used to measure quenching time-scales in nearby green valley galaxies; and 2) the resulting star formation quenching time-scales are longer for barred green valley galaxies. Considering that the presence of a bar indicates that more violent processes (e.g., major mergers) are absent in host galaxies, we conclude that the presence of a bar can be used as a morphological signature for slow star formation quenching.}


   \maketitle
%

\section{Introduction}

It is now well known that galaxies in the local Universe can generally be separated into two broad classifications with respect to their star formation activity: blue star-forming (\textit{blue cloud}) galaxies, composed mostly by disc galaxies, and red passive systems (\textit{red sequence}) typically composed by elliptical and lenticular galaxies with little or no recent star formation \citep[e.g., ][]{Strateva2001, Kauffmann2003, Shen2003, Baldry2004, Balogh2004, Menci2005}. The star formation rate (SFR)  of blue galaxies is well correlated with their stellar mass (M$_{*}$), defining the main-sequence of star-forming galaxies in the SFR-M$_{*}$ plane \citep{Brinchmann2004, Salim2007, Gilbank2010, McGee2011}. This correlation was already in place at $z\sim1$ \citep{Willmer2006, Noeske2007} and possibly at $z\geq2$ \citep[e.g., ][]{Brammer2009, Taylor2009, Whitaker2012, Speagle2014}. Most tellingly, the number density of red passive galaxies has roughly doubled  since  $z\sim1$, while  the  number  of  blue  galaxies has remained nearly constant \citep{Bell2004, Brown2007, Faber2007, Ilbert2013, Muzzin2013b, Sobral2014}. This motivates the search for processes that cause blue, star-forming galaxies to transition to red, non-star-forming ones. This transition has become known as \textit{quenching}.

Between the blue cloud and the red sequence peaks on galaxy colour-magnitude diagram (CMD) there is a narrow region, sparsely populated by galaxies, known as \text{the} \textit{green valley} \citep[][]{ Martin2007, Salim2007, Wyder2007, Schawinski2014, Renzini2015, Smethurst2015, Diaz-Garcia2017, Nogueira-Cavalcante2018}. The physical properties of galaxies in the green valley region are different from those in either the blue cloud or the red sequence, exhibiting intermediate characteristics, for instance the age of stellar population and local environment \citep{Pan2013}, the specific star formation rate \citep{Salim2014} and galaxy structures parametrized by S\'ersic indices \citep{Bremer2018}. The intermediate nature of green valley suggests that green valley galaxies are quenching their star formation, starting in the blue cloud phase, reaching the red sequence afterwards. The typical quenching time-scales, i.e., the duration of the green valley phase, is a function of cosmic time; the galaxy transition across the green valley in the local Universe ($z<0.2$) takes, on average, $\sim1-2$ Gyrs \citep{Martin2007, Bremer2018}, whereas at intermediate redshifts ($z\sim0.7$) the average of green valley quenching phase lasts less than 500 Myrs \citep{Goncalves2012}. 
External events can trigger new episodes of star formation in quiescent red galaxies, the galaxy bluewards through the CMD. \citet{Pan2014} found that more than 50\% of green valley early-type galaxies at $0.02<z<0.05$ have blue central cores, supporting a scenario of recent gas-rich mergers. This result is in agreement with simulations that predict that up to $\sim17\%$ of green valley galaxies at $z<2$ are passive galaxies increasing their star formation activity \citep{Trayford2016}. These galaxies could be early-type galaxies which live in the centres of galaxy clusters and occasionally capture a lower-mass gas-rich galaxy \citep[e.g., ][]{Martin2017, Darvish2018}. However, it is observed that the red sequence region has grown, at least, since $z\sim1$ \citep{Bell2004, Brown2007, Faber2007, Ilbert2013, Muzzin2013b} and, therefore, it is reasonable to consider that, for the majority of galaxies, the galaxy transition on CMD is from the blue cloud to the red sequence, passing through the green valley. 

Many possible quenching mechanisms that lead galaxies to the green valley have been proposed, for both low and high redshifts. Overall, these mechanisms can be classified as \textit{rapid processes}, i.e., those able to quench star formation on a scale of hundreds of Myrs, and \textit{slow processes}, taking more than 1 Gyr to completely quench star formation. Rapid processes can encompass, for instance, negative feedback from active galactic nuclei AGN \citep[e.g., ][]{DiMatteo2005, Springel2005a, Ciotti2007, Cattaneo2009, Martin2007, Schawinski2009, Dubois2013, Bongiorno2016, Smethurst2016, Kocevski2017}, major mergers \citep[e.g., ][]{Toomre1977, Springel2005d, Conselice2006} and violent disc instability \citep[e.g., ][]{Dekel2014, Zolotov2015, Nogueira-Cavalcante2019}. These mechanisms are claimed to explain the formation of most passive dead galaxies. Stellar feedback \citep[e.g., ][]{Menci2005, Lagos2013} is another example of a rapid process and is generally claimed to explain quenching in low-mass galaxies. Slow processes can involve interactions in clusters and groups \citep[e.g., ][]{Gunn1972, White1976, Hausman1978, Moore1998,  Abadi1999, Balogh1999, Balogh2004, Boselli2006, Capak2007, Moran2007, vandenBosch2008, Wetzel2013, Peng2015, Darvish2016, Hatfield2017, Smethurst2017, vandeVoort2017, Darvish2018} and secular evolution through internal galaxy structures like spiral arms and, specially, stellar bars \citep[e.g., ][]{Kormendy2004, Sheth2005, Masters2010, Masters2011, Mendez2011, Cheung2013}. Although different galaxy mechanisms can act simultaneously, observations suggest that quenching processes that correlate with stellar mass are independent from those that correlate with galaxy environment \citep{Peng2010, Sobral2014}.   

Even though the details concerning the physical processes responsible for the galaxy transition across the green valley are not known, recent works have given clues on how quenching processes act in galaxies. \citet{Rowlands2018} found that, at $z<1$, the red sequence has grown faster for intermediate- mass galaxies (M$_{\star}>10^{10.6}$ M$_{\odot}$) than for those with high stellar masses (M$_{\star}>10^{11}$ M$_{\odot}$). Furthermore, evidence for  multiple evolutionary quenching pathways has been shown by \citet{Schawinski2014} and \citet{Maltby2018} for galaxies at low and high redshifts, respectively. These results suggest that multiple quenching processes act simultaneously and in different ways at different cosmic times. Therefore, the green valley must present different characteristics in different epochs of the Universe, since, by definition, it is a transitional region of galaxies. In fact, the population of green valley galaxies is different in the local Universe, exhibiting mostly S0 and Sa disc morphologies \citep{Salim2014, Schawinski2014, Evans2018, Bremer2018}, as opposed to those at intermediate redshifts, which often exhibit elliptical and distorted morphologies \citep{Nogueira-Cavalcante2018}. These general results suggest that violent processes (associated with more elliptical and distorted morphologies at intermediate redshifts) are more frequent for a younger Universe whereas secular evolution is more common at low redshifts.   

A useful way to understand how galaxies transition from the blue cloud to the red sequence is by measuring how long it takes for galaxies to cross the green valley. \citet{Martin2007} developed a method that estimates quenching time-scales for green valley galaxies (Section \ref{measuring_star_formation_quenching_time_scales}), based on photometry and spectroscopy from Sloan Digital Sky Survey \citep[SDSS, ][]{York2000} and photometry from Galaxy Evolution Explorer \citep[GALEX, ][]{Martin2005}. However, one difficulty with this method is the need spectroscopic data. Big spectroscopic surveys, like SDSS, zCOSMOS \citep{Lilly2007} and LEGA-C \citep{Wel2016, Straatman2018} pre-select only the brightest galaxies. Moreover, in the case of SDSS galaxy spectra, the fiber diameter is 3$^{\prime\prime}$, restricting observations to the centre of nearby galaxies.   

We develop in this work a new approach to estimate quenching time-scales in green valley galaxies that circumvents this difficulty, based on the \cite{Martin2007} method, but using only photometric galaxy data. The lack of spectra is compensated by the narrow band filters of the photometric system of the Javalambre Photometric of Local Universe Survey\footnote{www.j-plus.es} \citep[J-PLUS,][]{Cenarro2019}. This is an ancillary survey of the Javalambre Physics of the accelerating universe Astrophysical Survey\footnote{www.j-pas.org} \citep[J-PAS, ][]{Benitez2014}, which will observe 8500 deg$^2$ of the sky with 56 narrow band filters. J-PLUS observations are carried out from the Javalambre Observatory in Spain, with the Javalambre Auxiliary Survey Telescope (JAST/T80), which is an 0.83m telescope, equipped with the camera T80Cam, supplying a field of view of 3 deg$^2$ in a pixel-scale of 0.55". Additionally, its filter set is composed by a total of 12 optical filters: 4 broad ($g$, $r$, $i$ and $z$), 2 intermediate ($u$JAVA and F0861) and 6 narrow (F0378, F0395, F0410, F0430, F0515 and F0660).

In this work we show the application of this new approach in a sample of nearby green valley disc galaxies and we analyze the dependence of the measured quenching time-scales of these galaxies on their probability of hosting a bar. Stellar bars are present in more than half of nearby spiral galaxies \citep[e.g., ][]{Menendez-Delmestre2007, Sheth2008, Masters2011} and are associated with the redistribution of angular momentum of the baryonic and dark matter components \citep{Sellwood1981, Athanassoula2003, Athanassoula2013, Machado2010}, formation of rings and lens structures \citep{Kormendy2004}, induction of star formation activity in the central parts of host galaxies \citep{Sheth2005, Coelho2011, Oh2012, Wang2012}, formation of pseudobulges \citep{Sheth2005, Mendez-Abreu2008, Aguerri2009}, and possibly feeding active galactic nucleus (AGN) activity \citep{Shlosman1989, Ho1997, Oh2012}. 

We structure this paper as follows. In Section~\ref{measuring_star_formation_quenching_time_scales} we describe the original method by \citet{Martin2007} to measure quenching time-scales and our new approach based on that method. In Section~\ref{quenching_time_scales_as_a_function_of_host_a_bar} we apply our method to nearby green valley disc galaxies and analyzed how their morphological parameters can impact quenching time-scales. In Section~\ref{discussions_and_conclusions} we discuss the robustness of our approach. The results are summarized in Section \ref{summary}. Throughout this paper we adopt AB magnitudes and standard $\Lambda$CDM cosmology: $H_0=70$ km s$^{-1}$ Mpc$^{-1}$, $\Omega_{\text{m}}=0.30$ and $\Omega_{\Lambda}=0.70$.


\section{Measuring star-formation quenching time-scales in green valley galaxies}\label{measuring_star_formation_quenching_time_scales}

In this work we develop a method that estimates quenching time-scales in nearby green valley galaxies, with J-PLUS and GALEX photometric data, and based on the work described in \citet{Martin2007}. The main difficulty in that approach with J-PLUS data is that we cannot directly compare the J-PLUS magnitudes with SDSS spectral indices due to the limited fiber aperture of $3^{\prime\prime}$ in diameter of SDSS spectra. For instance, up to $z\sim 0.2$, only the bulge of a Milky Way-like galaxy is observed with such aperture. One way to circumvent this issue is through constructing J-PLUS spectral energy distributions (SED) from SDSS spectra. To that end, we used a \texttt{python} package called \texttt{pysynphot} \citep{pysynphotguide, pysynphotcode} as well as the appropriate parameters given the survey/telescope properties \citep[for further details we refer the reader to][]{Cenarro2019}. In next sections we describe the selection of green valley galaxies in SDSS data, the original \citet{Martin2007} method that this work is based on and our new approach that measure quenching time-scales in green valley galaxies from J-PLUS photometry.

\subsection{Selection of green valley galaxies}\label{selection_of_green_valley}

\begin{figure}
\includegraphics[width=\columnwidth]{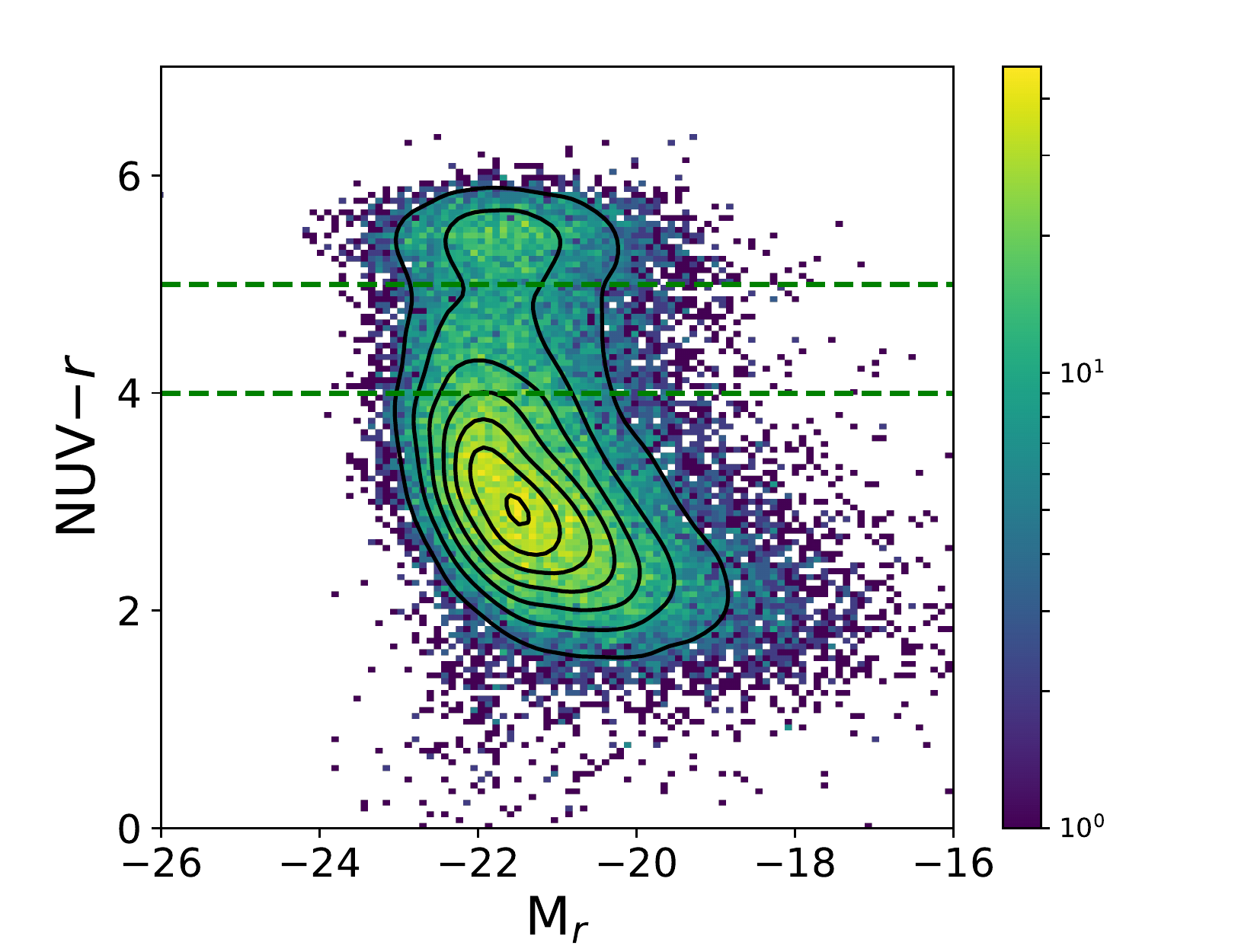}
\includegraphics[width=\columnwidth]{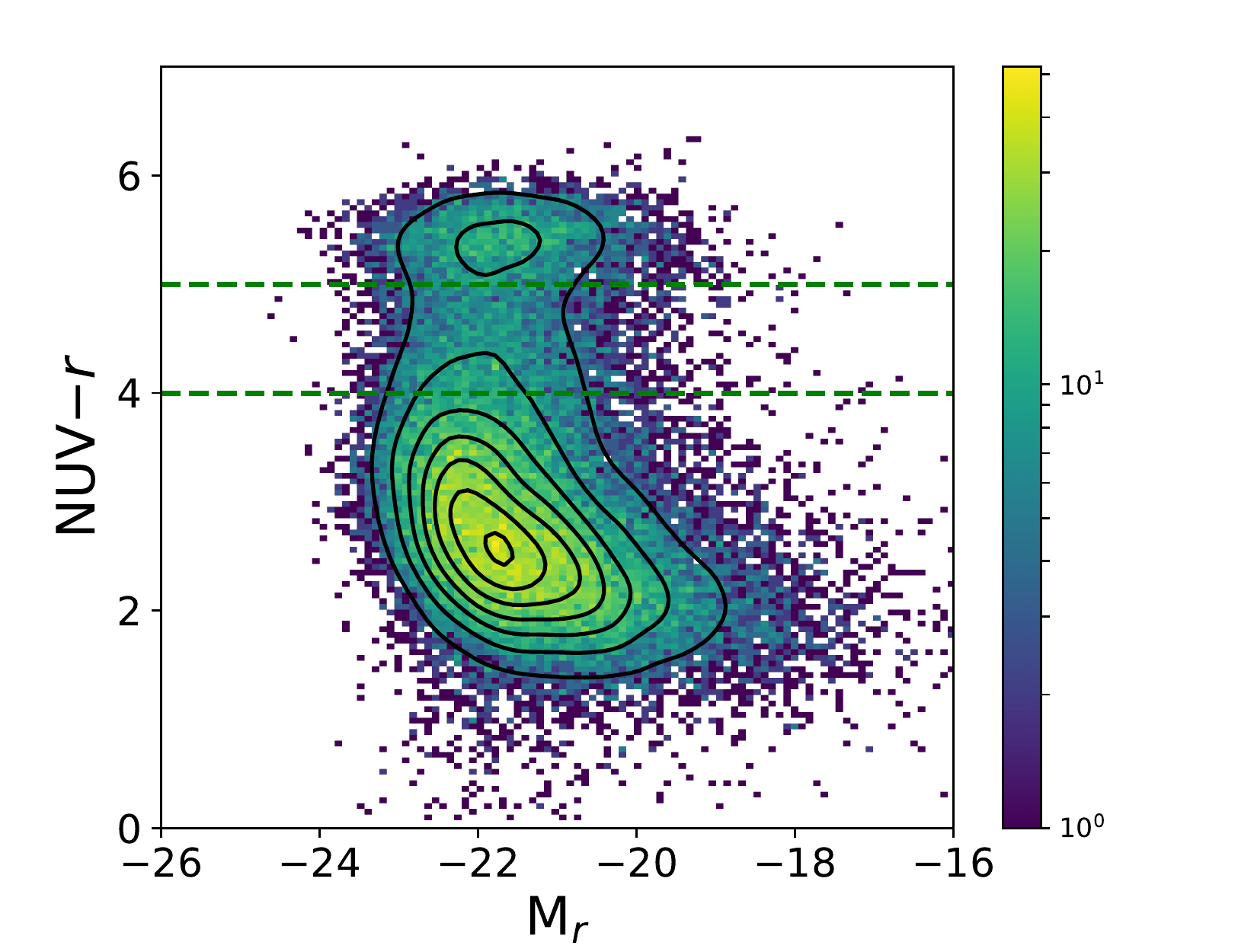}
\caption{Colour-magnitude diagrams of SDSS+GALEX galaxies at $z<0.2$, uncorrected (top panel) and corrected (bottom panel) by intrinsic extinction caused by dust. The horizontal dashed lines represent the green valley region. Each contour line represents approximately the same number of galaxies. The colours in the vertical bar on both panels identify the number of galaxies for each bin within the diagrams.}
\label{colour_magnitude_diagram}
\end{figure}

We select green valley galaxies from the traditional galaxy CMD (Fig. \ref{colour_magnitude_diagram}), using two broad bands: the Near-Ultraviolet (NUV) band, from GALEX GR6/GR7 \citep{Martin2005}, and $r$ band, from SDSS DR12 \citep{Alam2015}. We retrieve spectra and photometry through GALEX CasJobs database\footnote{https://galex.stsci.edu/casjobs/0}; also, following the selection criteria of \citet{Martin2007}: redshift range of~$0.0<z<0.2$; NUV weight~$>800$; NUV artifact~$< 2$; apparent NUV magnitude~$< 23$; and apparent $r$ magnitude~$< 17.7$. We cross-match SDSS and GALEX samples within a radius of $1^{\prime\prime}$  and apply \citet{Seaton1979} model to correct for Galactic extinction and k-correct SDSS and GALEX magnitudes using the k-correction calculator \citep[the \texttt{python} version of the K-correction calculator\footnote{http://kcor.sai.msu.ru/getthecode/}, ][]{Chilingarian2010, Chilingarian2012}). We use the definition of the green valley on the CMD as the region between $4.0<\text{NUV}-r<5.0$, as estimated by previous works \citep[e.g., ][]{Martin2007, Goncalves2012, Salim2014, Nogueira-Cavalcante2018, Nogueira-Cavalcante2019}. From the green valley sample we sub-select those galaxies with error of H$_{\delta,A}$ spectral index lower than 1 $\mbox{\AA}$. This is a necessary condition to properly determine the best star formation history model from stellar population synthesis codes \citep[see Section \ref{Martin2007_method} and also ][]{Goncalves2012}. With these constraints, the initial number of green valley galaxies is comprised of 2,865 objects.

The green valley is not only populated by transitional galaxies but also by star-forming galaxies obscured by dust. We properly correct galaxy magnitudes through the flux of the H$_{\alpha}$ and H$_{\beta}$ emission lines, estimating the excess in $B-V$ colour, $E(B-V)$, following the prescription by  \citet{Calzetti1994}:

\begin{equation}
E(B-V) = 0.935 \ln\left(\frac{\text{H}_{\alpha}/\text{H}_{\beta}}{2.88}\right) \times 0.44 \,\, ,
\end{equation}
where the term 0.44 takes into account that we are estimating stellar continuum extinction through nebular emission \citep{Calzetti1994}. In other words, the extinction in the stellar continuum is roughly half of the value for ionized gas. To convert $E(B-V)$ into extinction in SDSS and GALEX magnitudes we use the relation:
\begin{equation}
A_{\lambda} = k_{\lambda} \times E(B-V) \,\, ,
\end{equation}
where $A_{\lambda}$ is the extinction in magnitudes at the wavelength $\lambda$ and $k_{\lambda}$ is the extinction curve by \citet{Calzetti2000}. For $0.12 \leq \lambda[\mu\text{m}] \leq 0.63$:
\begin{equation}
k(\lambda)=1.17\left(-2.156+\frac{1.509}{\lambda}-\frac{0.198}{\lambda^2} + \frac{0.011}{\lambda^3}\right)+1.78;
\end{equation}
and for $0.63 \leq \lambda[\mu\text{m}] \leq 2.20$:
\begin{equation}
k(\lambda) = 1.17\left(-1.857+\frac{1.040}{\lambda}\right)+1.78.
\end{equation}
 
The CMD corrected by intrinsic dust extinction is shown in Fig. \ref{colour_magnitude_diagram}. The final number of green valley galaxies 2,043, after excluding star-forming galaxies obscured by dust from the initial sample.

\subsection{\citet{Martin2007} method}\label{Martin2007_method}

\begin{figure}
\includegraphics[width=\columnwidth]{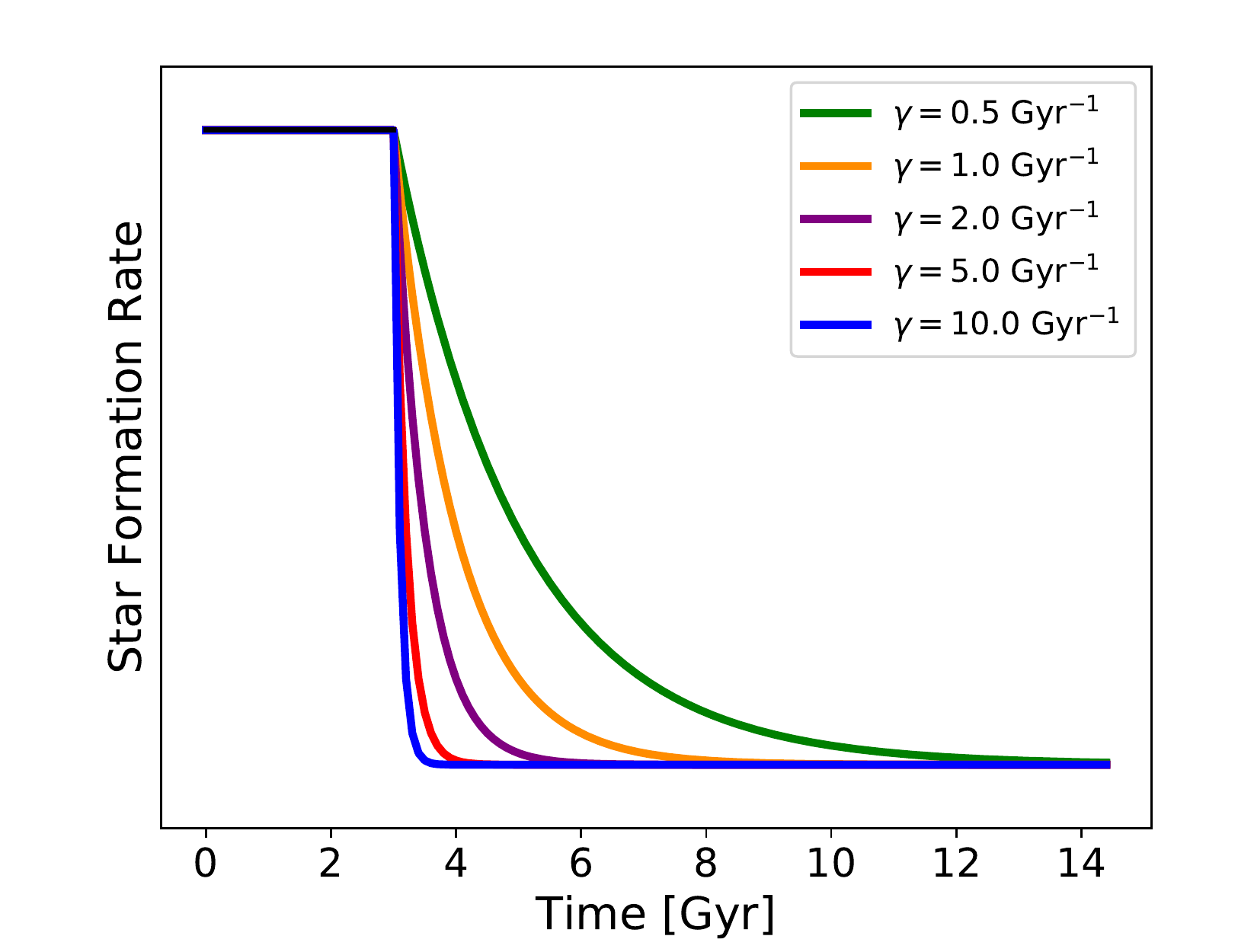}
\caption{Star formation rate models for five different $\gamma$ values considered in this work (Eqs. \ref{star_formation_history_1} and \ref{star_formation_history_2}).}
\label{star_formation_history_model}
\end{figure}

\begin{figure*}

\includegraphics[width=6.3cm]{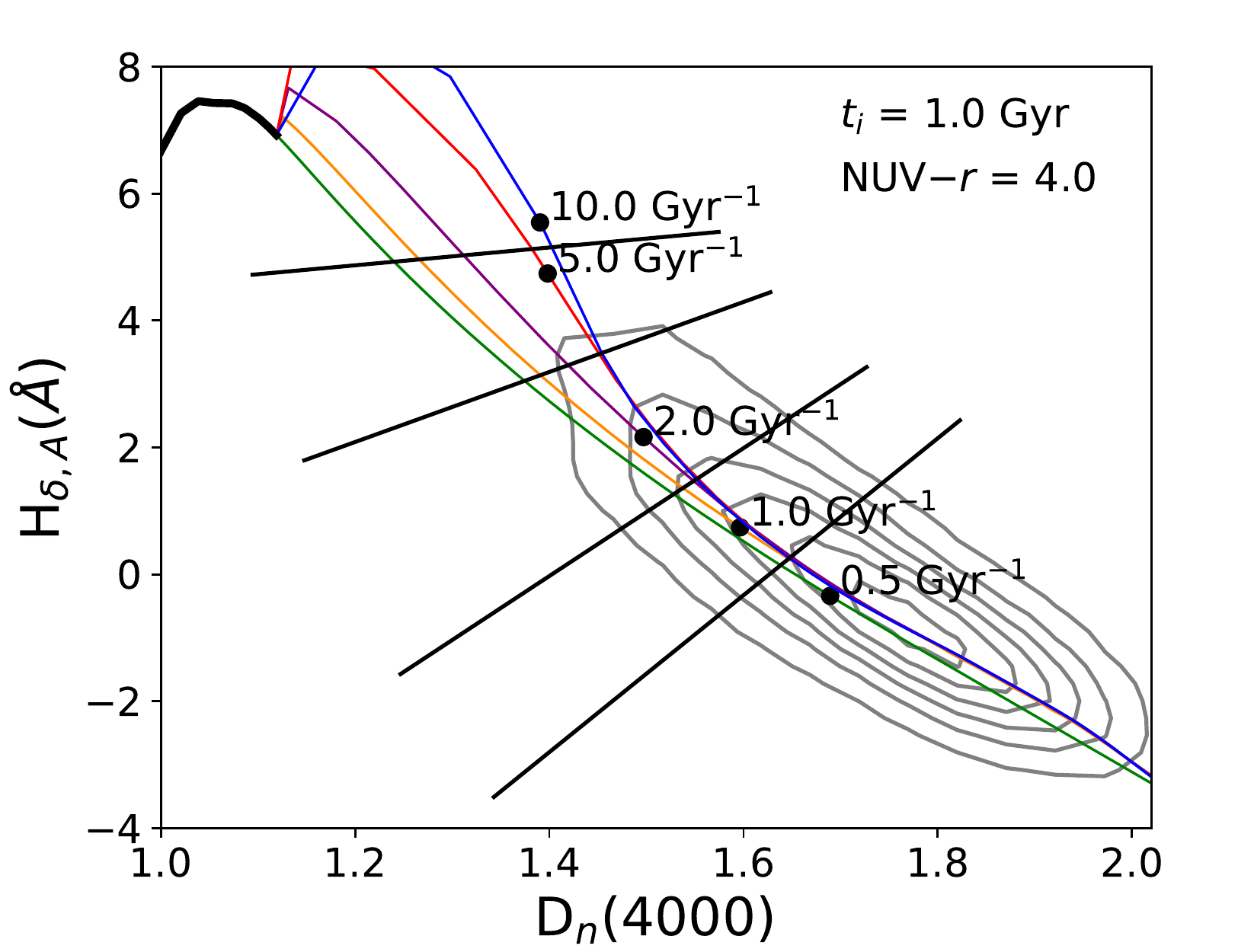}
\includegraphics[width=6.3cm]{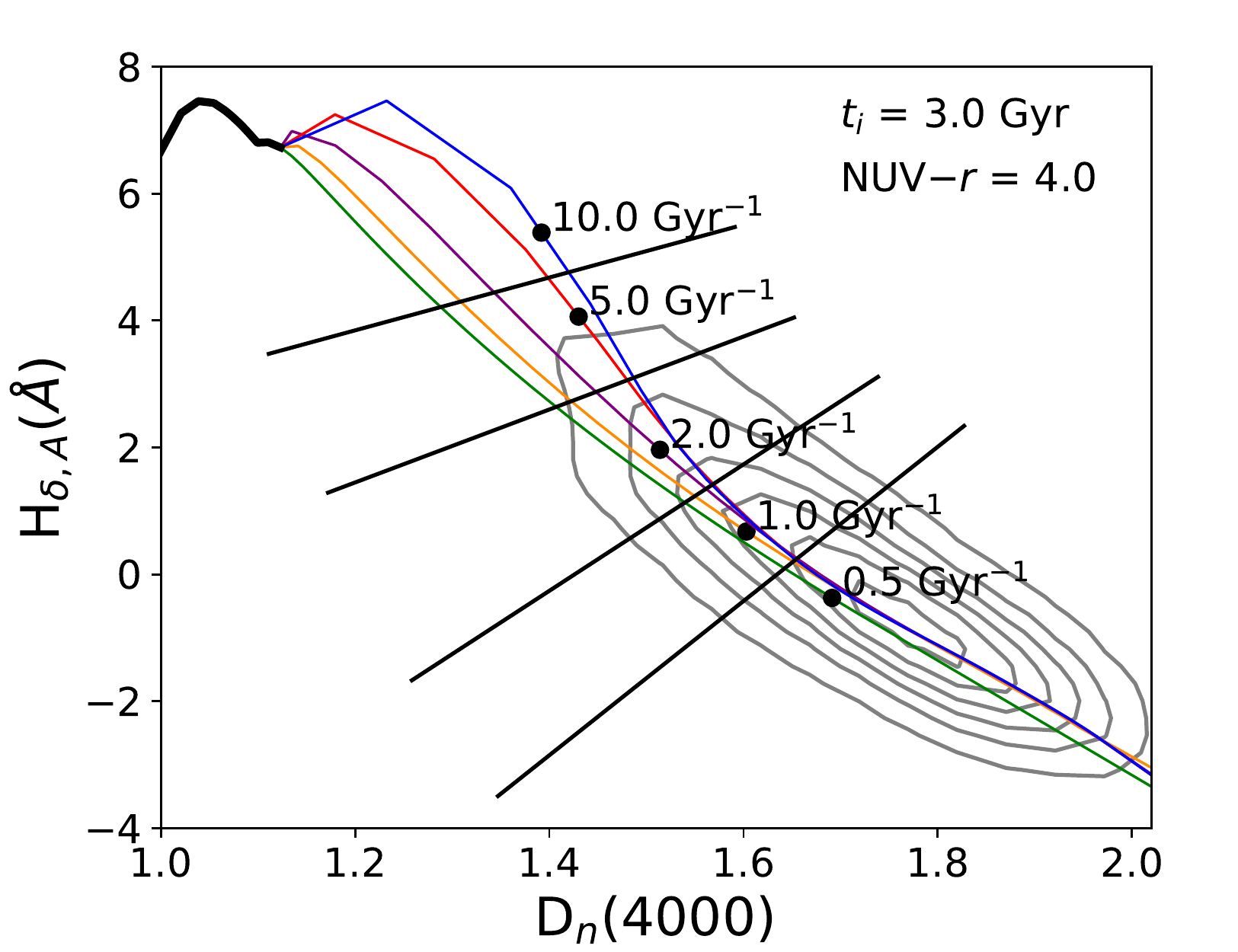}
\includegraphics[width=6.3cm]{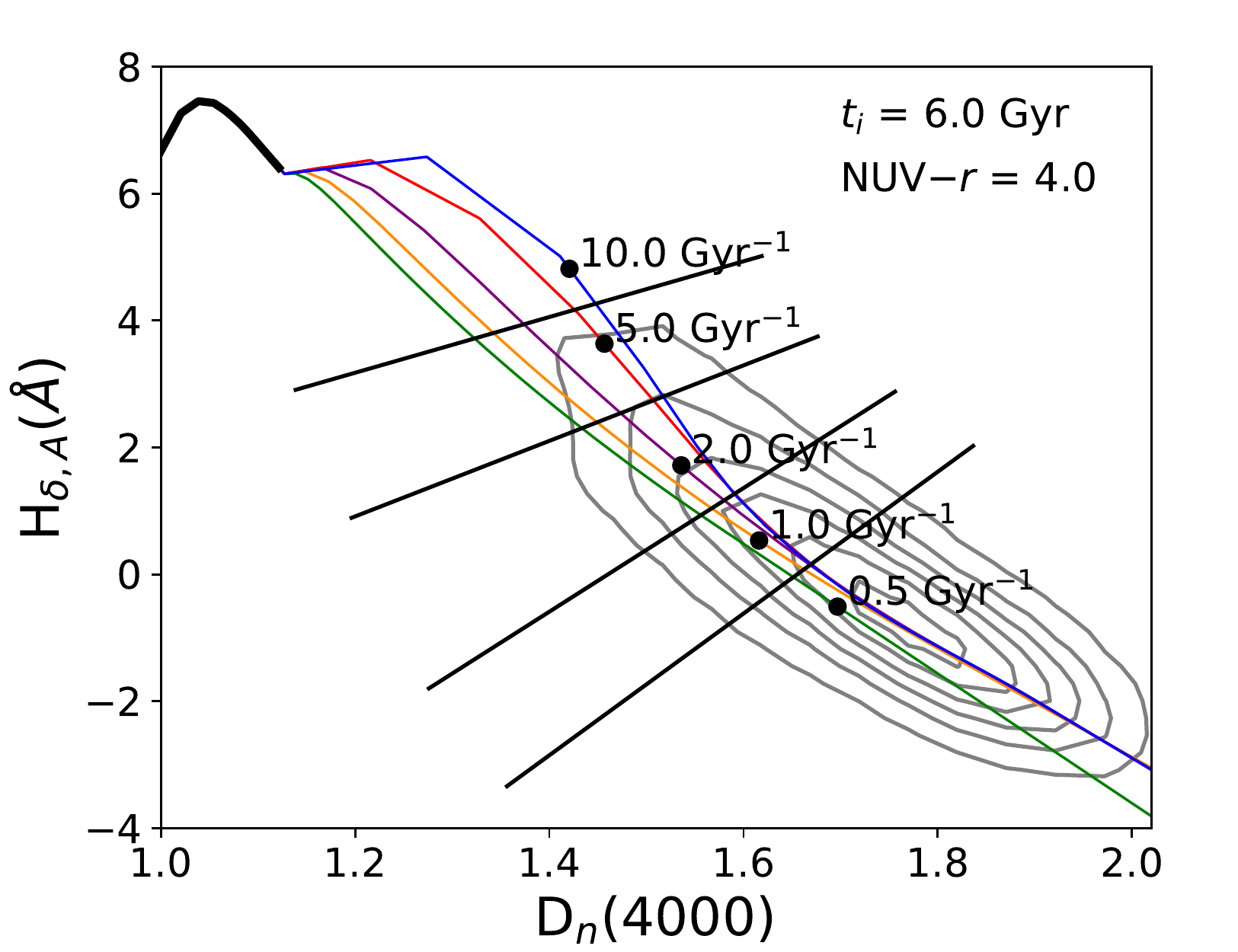}

\includegraphics[width=6.3cm]{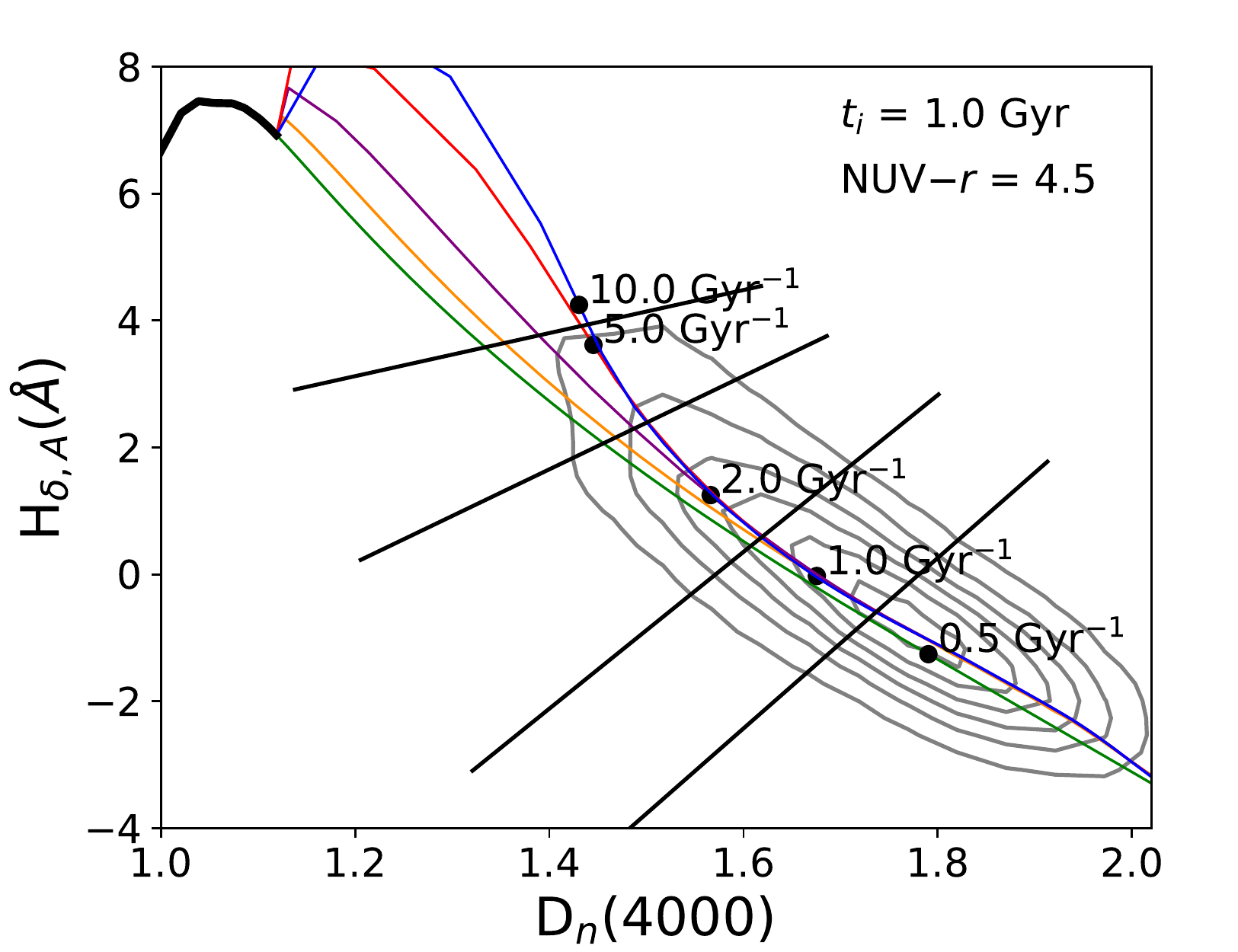}
\includegraphics[width=6.3cm]{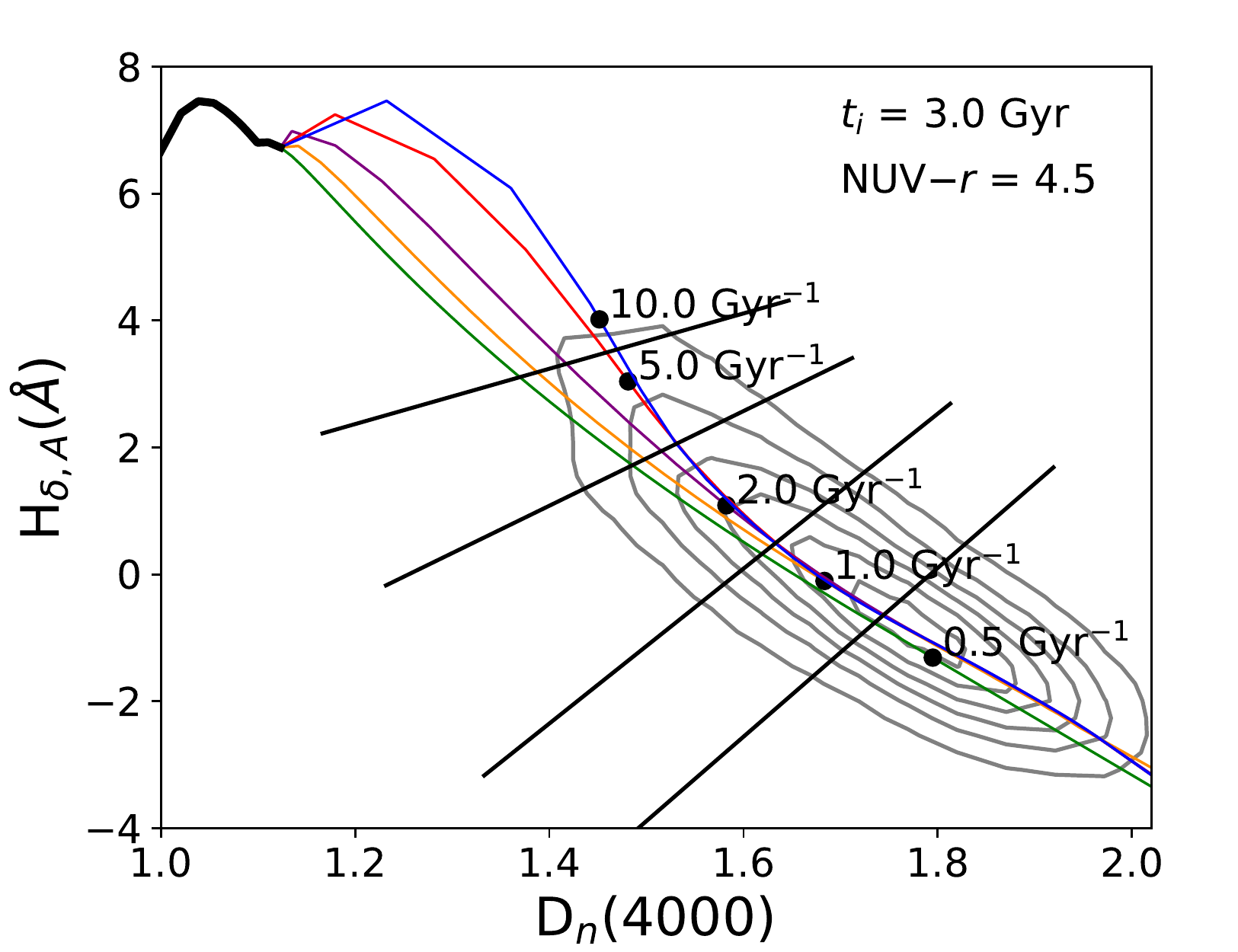}
\includegraphics[width=6.3cm]{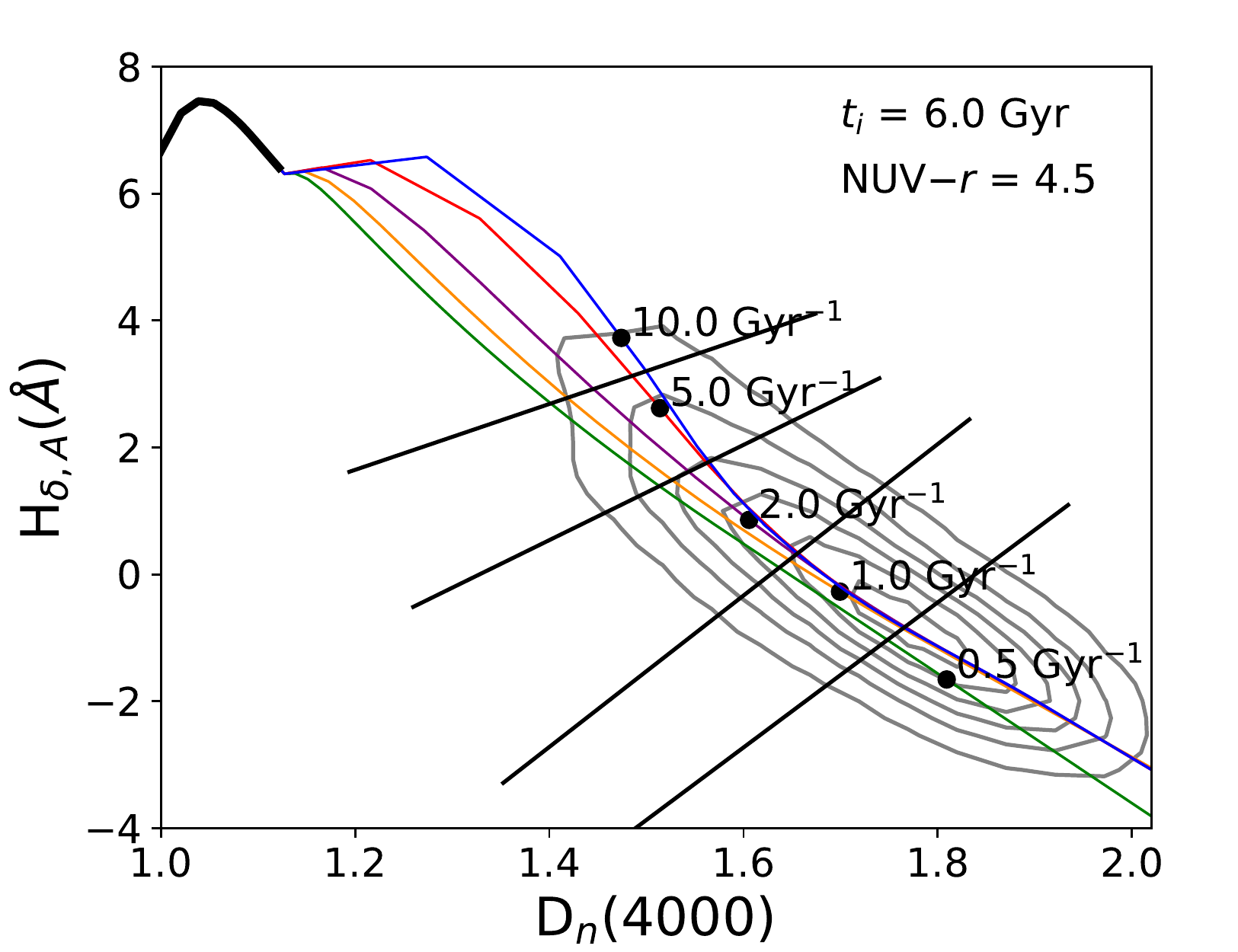}
\caption{$\text{H}_{\delta,A} \times \text{D}_{n}(4000)$ planes, considering five $\gamma$ values from the SFH model described in Eqs. \ref{star_formation_history_1} and \ref{star_formation_history_2} (0.5 Gyr$^{-1}$, 1 Gyr$^{-1}$, 2 Gyr$^{-1}$, 5 Gyr$^{-1}$ and 10 Gyr$^{-1}$). The black dots represent the spectral indices on the planes for a specific SFH and NUV$-r$ colour. Different NUV$-r$ colours show different spectral indices for the same SFH. The gray contours are the distribution of the spectral indices of green valley galaxies measured from SDSS spectra. The straight lines are a geometric average of the spectroscopic indices of} two consecutive black dots, dividing the plane in parts and allowing us to computationally interpolate it. For different values of $t_i$ the position of black dots does not change considerably, demonstrating that our model is weakly depend of the $t_i$ parameter.
\label{Distribution_green_valley_galaxies_on_HdA_vs_Dn4000_plane}
\end{figure*}

This method considers a very simple parametric model for the star formation histories (SFHs) of green valley galaxies (Fig. \ref{star_formation_history_model}): a constant star formation rate for the first billions of years, which corresponds to the blue cloud phase, followed by a period of exponential decay (green valley phase):

\begin{equation}
  \text{SFR}(t)=\text{SFR}(t_i), \,\,\,\,\ \text{for } t<t_i ;
  \label{star_formation_history_1}
 \end{equation}
 \begin{equation}
  \text{SFR}(t)=\text{SFR}(t_i)e^{-\gamma (t-t_i)}, \,\,\,\,\ \text{for } t>t_i \,\, ,
  \label{star_formation_history_2}
\end{equation}
where $t_i$ is a characteristic time and the exponential index ($\gamma$) is in units of Gyr$^{-1}$. This model is not strictly sensitive to $t_i$ (see Fig. \ref{Distribution_green_valley_galaxies_on_HdA_vs_Dn4000_plane}), since the parameters used to estimate $\gamma$ evolve very weakly during the star-forming phase \citep[for full details see ][]{Nogueira-Cavalcante2018}. For instance, \citet{Martin2007} used $t_i=10$ Gyrs, whereas \citet{Goncalves2012} and \citet{Nogueira-Cavalcante2018, Nogueira-Cavalcante2019} used $t_i=6$ Gyrs. In this work we consider $t_i=3$ Gyrs, which corresponds to the peak of star formation density of the Universe \citep[e.g, ][]{Madau2014}. It is worth mentioning that the only constraint is that $t_i$ must be greater than zero. The $\gamma$ index is the free parameter of this method, which reflects to the speed of quenching (for smaller $\gamma$, slower quenching; larger $\gamma$, faster quenching). The inverse of the exponential index (1/$\gamma$) corresponds to a time-scale, during which the star formation rate decreases $\sim37\%$ from the initial star formation rate before quenching phase ($e$-folding time). We call this time-scale hereafter as \textit{quenching time-scale}. 

To estimate $\gamma$ indices for green valley galaxies, \citet{Martin2007} consider three different observables, measured in the rest-frame: one photometric, the NUV$-r$, and two spectroscopic, the 4000 $\mbox{\AA}$ break \citep[D$_n$(4000),][]{Bruzual1983} and H$_{\delta}$ absorption line \citep[H$_{\delta,A}$,][]{Worthey1997}. These indices are defined as follows: 

\begin{equation}
  \text{D}_n(4000) = \sum_{\lambda = 4000 \text{\AA}}^{4100 \text{\AA}} F_{\lambda} \left/ \sum_{\lambda=3850 \text{\AA}}^{3950\text{\AA}} F_{\lambda} \right. 
 \label{Dn_4000_indice}
\end{equation} 
and
\begin{equation}
 \text{H}_{\delta,A} = \sum_{\lambda=4083,5 \text{\AA}}^{4122,25 \text{\AA}}\left(1 - \frac{F_{\lambda}}{F_{\lambda,\text{cont}}}\right) d\lambda \,\, ,
 \label{H_delta_A_indice}
\end{equation}
where $F_{\lambda,\text{cont}}$ is the continuum flux, defined by a straight line through the average flux density (in \AA) between 4041.60$-$4079.75 and 4128.50$-$4161.00.

Thus, $\gamma$ index is a function of these parameters, i.e., $\gamma(\text{NUV}-r, \text{D}_{n}(4000), \text{H}_{\delta,A})$. Particularly, the D$_{n}$(4000) and H$_{\delta,A}$ spectral indices define a well established plane, where each position indicates a different galaxy star formation history \citep{Kauffmann2003}. Fig. \ref{Distribution_green_valley_galaxies_on_HdA_vs_Dn4000_plane} shows a $\text{H}_{\delta,A} \times \text{D}_{n}(4000)$ plane, where coloured curves represent the expected temporal evolution of the values of these spectral indices from a stellar population synthesis model, considering the star formation history described by Eqs. \ref{star_formation_history_1} and \ref{star_formation_history_2}. We use five different $\gamma$ values to reproduce these curves: 0.5 Gyr$^{-1}$, 1 Gyr$^{-1}$, 2 Gyr$^{-1}$, 5 Gyr$^{-1}$ and 10 Gyr$^{-1}$, using \citet{Bruzual2003} models, with solar metallicities, Padova 1994 stellar evolutionary tracks \citep{Alongi1993, Bressan1993, Fagotto1994a, Fagotto1994b} and initial mass function from \citet{Chabrier2003}. The black dots for each colour lines on these planes represent the exact predicted value from \citet{Bruzual2003} models for a specific NUV$-r$ colour. In other words, for different NUV$-r$ colours there are different values of the spectral indices for the same SFHs. We also show in Fig. \ref{Distribution_green_valley_galaxies_on_HdA_vs_Dn4000_plane} the $\text{H}_{\delta,A} \times \text{D}_{n}(4000)$ plane for different values of $t_i$ (1 Gyr, 3 Gyr and 6 Gyr). The position of the black dots does not change considerably for different values of $t_i$, demonstrating that our model is weakly depend of the $t_i$ parameter. 

However, the SFHs of real green valley galaxies are more complex than those described by Eqs. \ref{star_formation_history_1} and \ref{star_formation_history_2}, reflecting a dispersed distribution of green valley spectral indices on $\text{H}_{\delta,A} \times \text{D}_{n}(4000)$ plane (gray curves in Fig. \ref{Distribution_green_valley_galaxies_on_HdA_vs_Dn4000_plane}). Therefore, we expect that the spectral indices measured from green valley galaxy spectra are not exactly those predicted from \citet{Bruzual2003} models. To take this into account the $\text{H}_{\delta,A} \times \text{D}_{n}(4000)$ plane is divided in parts, so that between two consecutive points (dots) we  define a perpendicular bisector, which corresponds to the geometric average of the two values. For instance, between the dots which correspond to 0.5 Gyr$^{-1}$ and 1 Gyr$^{-1}$ the bisector corresponds to $\sim$0.707 Gyr$^{-1}$. We interpolate the entire $\text{H}_{\delta,A} \times \text{D}_{n}(4000)$ plane, associating each coordinate to a single specific $\gamma$ value. For more details on this process see \citet{Nogueira-Cavalcante2018}. 

\subsection{New photometry-only approach with J-PLUS}

\begin{figure}
\includegraphics[width=\columnwidth]{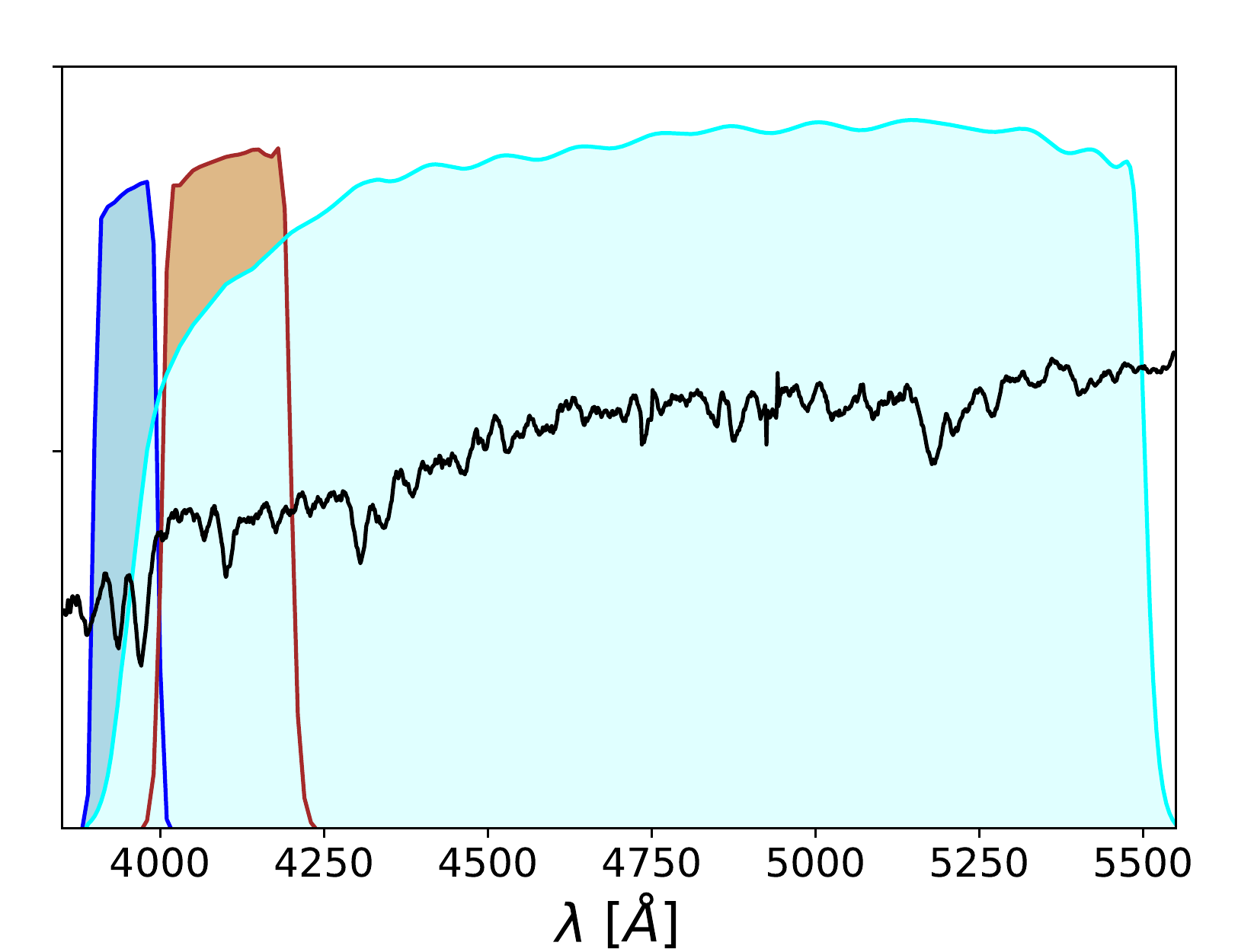}
\caption{Example of a SDSS green valley galaxy spectrum with three J-PLUS filter responses (F0395 in blue, F0410 in brown and gSDSS in cyan). Note that the F0395 and F0410 filters straddle the 4000 $\AA$ break region, whereas the H$_{\delta,A}$ absorption line (4102 $\AA$) is right in the middle of the F0410 filter.}
\label{filters_response_J-PLUS_with_a_galaxy_spectrum}
\end{figure}

\begin{figure}
\includegraphics[width=\columnwidth]{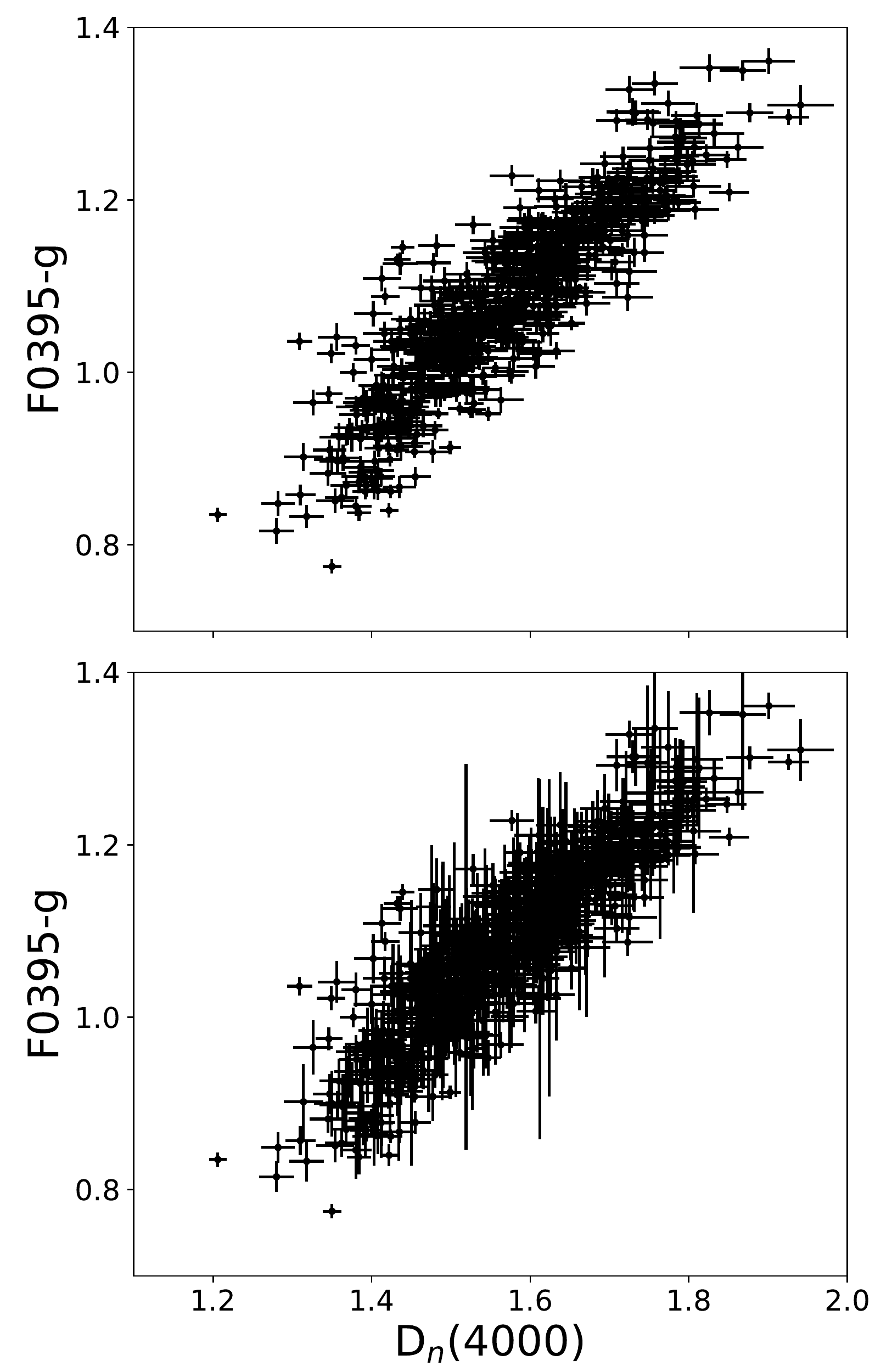}
\caption{F0395$-g$ colour as a function of D$_{n}$(4000) spectral index. The top panel does not take into account internal extinction whereas the bottom panel is corrected by dust reddening.}
\label{F0395_g_as_a_function_of_Dn4000}
\end{figure}

\begin{figure}
\includegraphics[width=\columnwidth]{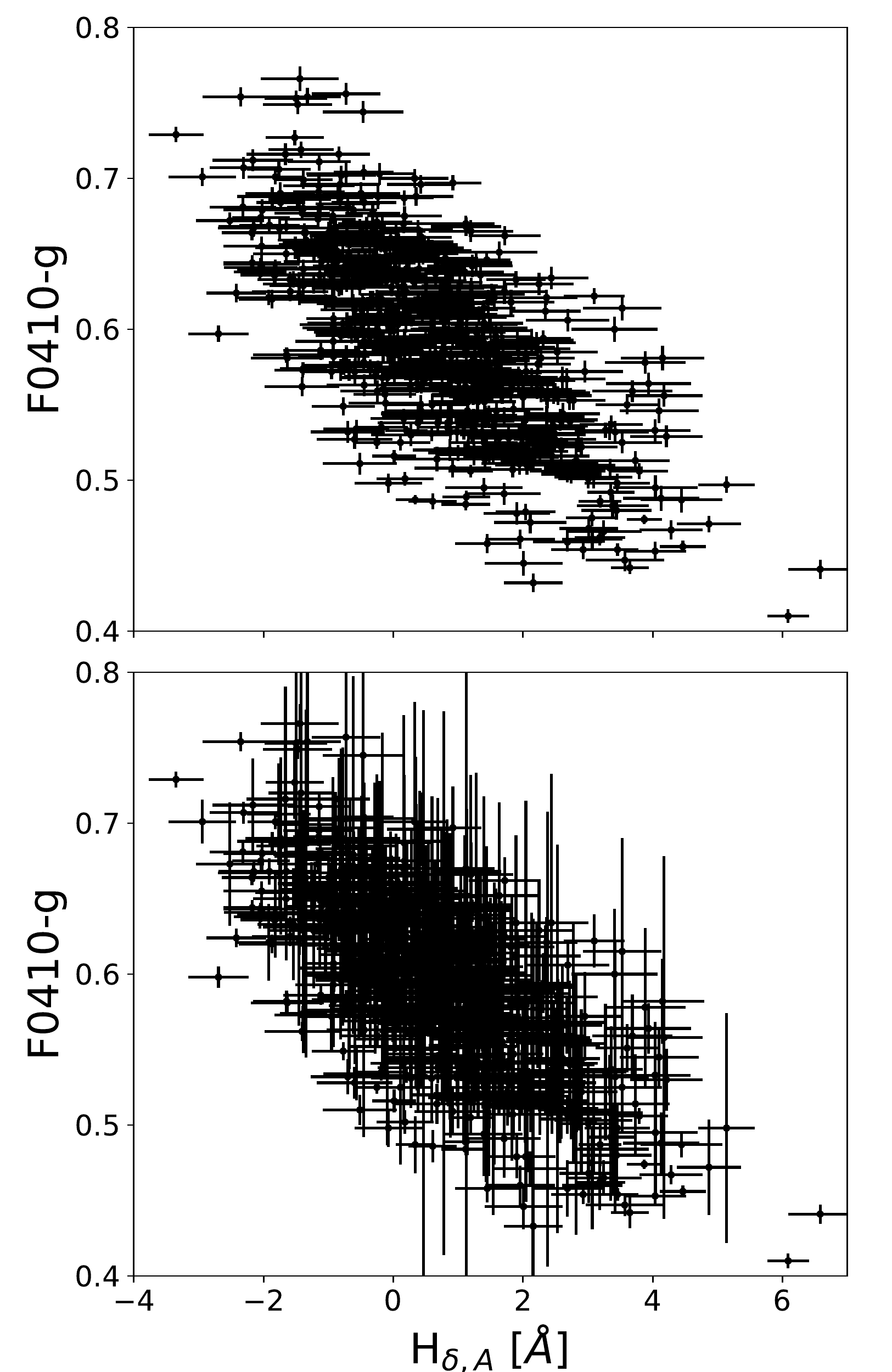}
\caption{F0410$-g$ colour as a function of H$_{\delta,A}$ spectral index, where the F0410$-g$ colour is corrected (top panel) and non-corrected (bottom panel) by dust extinction.}
\label{F0410_g_as_a_function_of_HdA}
\end{figure}

We seek a set of J-PLUS filters that can, efficiently, substitute the D$_{n}$(4000) and H$_{\delta,A}$ spectral indices. Therefore, for each SDSS spectrum we estimate a J-PLUS SED in the F0395, F0410 and $g$ J-PLUS filters. We chose these filters because they lie approximately in the same region of the D$_n$(4000) and H$_{\delta, A}$ spectral indices, as shown in Fig. \ref{filters_response_J-PLUS_with_a_galaxy_spectrum}. Figs. \ref{F0395_g_as_a_function_of_Dn4000} and \ref{F0410_g_as_a_function_of_HdA} show the F0395$-g$ and F0410$-g$ colours as functions of D$_n$(4000) and H$_{\delta, A}$ spectral indices, respectively. There are clear correlations among these quantities, demonstrating that, in principle, these colours can be used instead of the D$_n$(4000) and H$_{\delta, A}$ spectral indices to estimate quenching time-scales.

\begin{figure}
\includegraphics[width=\columnwidth]{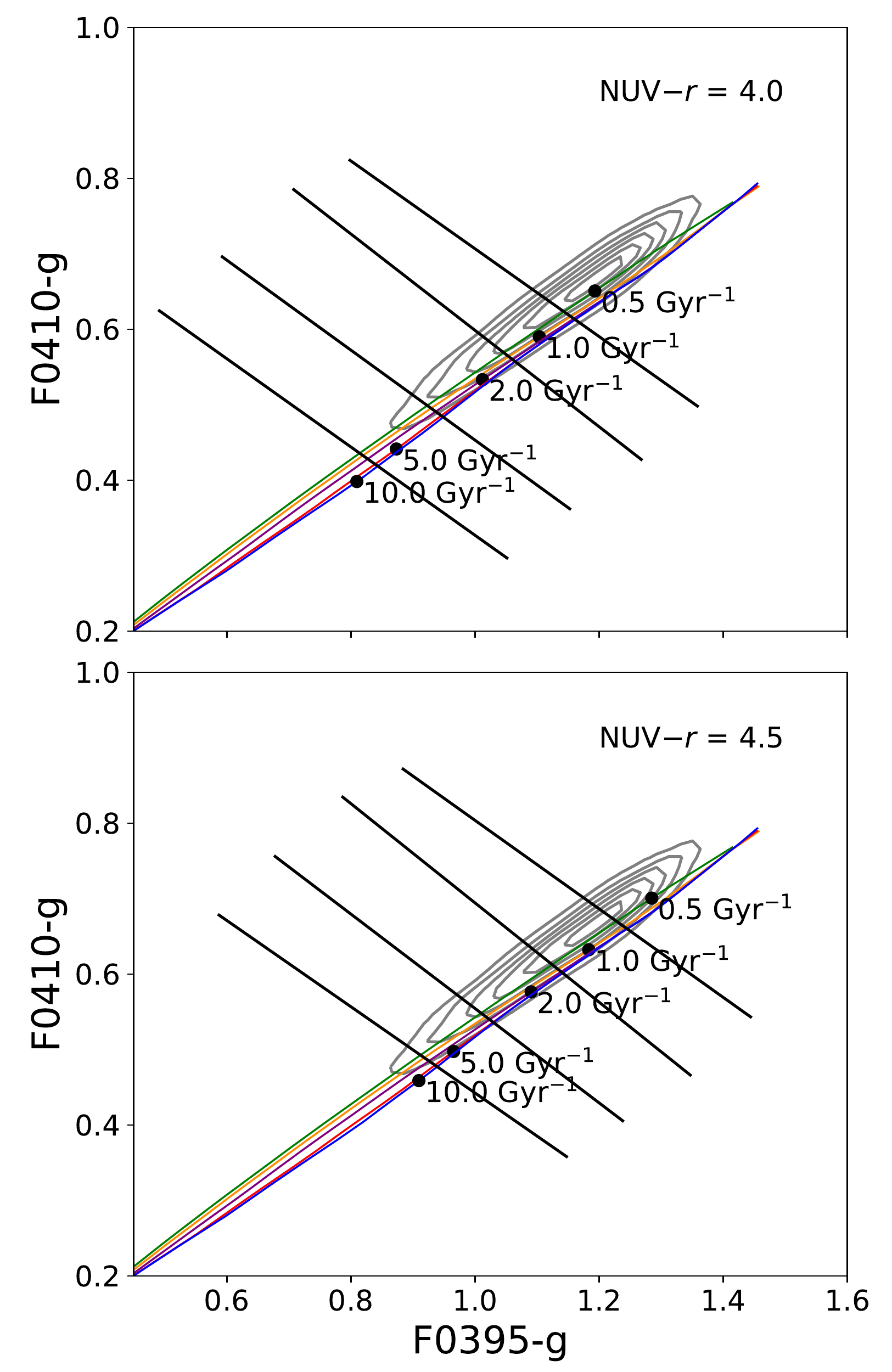}
\caption{$\text{F0410}-g \times \text{F0395}-g$ colour-colour diagrams, showing in coloured lines the SFH models presented in Eqs. \ref{star_formation_history_1} and \ref{star_formation_history_2} with different $\gamma$ values. The black dots and the gray contours are the same as in Fig. \ref{Distribution_green_valley_galaxies_on_HdA_vs_Dn4000_plane}, for a fixed NUV$-r$ colour}.
\label{Distribution_green_valley_galaxies_on_F0395_g_vs_F0410_g_plane}
\end{figure}

\begin{figure}
\includegraphics[width=\columnwidth]{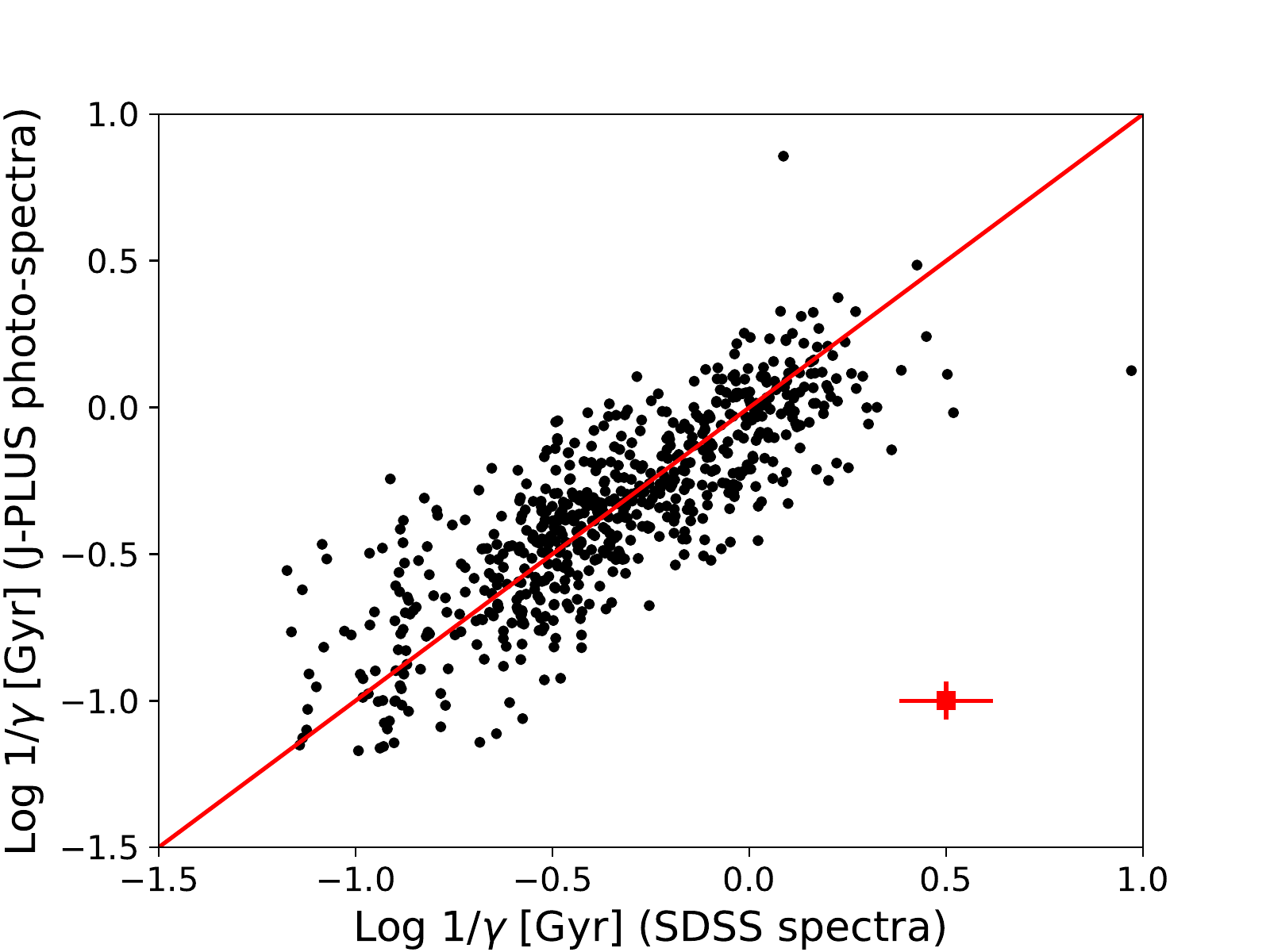}
\caption{Quenching time-scales in green valley galaxies from J-PLUS SEDs versus quenching time-scales from spectral indices, for galaxies whose colours are corrected by dust reddening. The red square merely represents the average uncertainties in our} estimates. The spearman correlation and the root mean squared error (RMSE) are, respectively, 0.84 and 0.2, indicating a good correlation between the variables.
\label{Quenching_timescales_from_JPLUS_photometry_and_SDSS_spectra_with_intrinsic_ext_corr}
\end{figure}

Fig. \ref{Distribution_green_valley_galaxies_on_F0395_g_vs_F0410_g_plane} shows the same distribution of green valley galaxies and the predicted curves by \citet{Bruzual2003} models as in Fig. \ref{Distribution_green_valley_galaxies_on_HdA_vs_Dn4000_plane}, but now on the $\text{F0410}-g \times \text{F0395}-g$ plane. The procedure to estimate $\gamma$ indices from this diagram is similar to that followed with the $\text{H}_{\delta,A} \times \text{D}_{n}(4000)$ plane. For each two consecutive points, which correspond to the predicted values by \citet{Bruzual2003} models for a specific NUV$-r$ value, we construct a bisector, corresponding to the geometric average of the involved dots. Following this procedure we divided all the $\text{F0410}-g \times \text{F0395}-g$ plane in parts, similarly to the $\text{H}_{\delta,A} \times \text{D}_{n}(4000)$ plane, by \citet{Martin2007}. We then compared the estimated quenching time-scales between \citet{Martin2007} and our new approach (Fig. \ref{Quenching_timescales_from_JPLUS_photometry_and_SDSS_spectra_with_intrinsic_ext_corr}).

\section{Star formation quenching time-scales in nearby green valley disc galaxies and constraints with bars}\label{quenching_time_scales_as_a_function_of_host_a_bar}

By cross-matching the green valley galaxies selected from Fig. \ref{colour_magnitude_diagram} with J-PLUS Data Release 1\footnote{www.j-plus.es/datareleases/data\_release\_dr1} (DR1) and the morphological catalog from \citet{DominguezSanchez2018}, we identify which galaxies in our sample were observed by J-PLUS and characterized their probability of hosting a bar. In the following, we describe \citet{DominguezSanchez2018} catalog and our results, associating quenching time-scales of green valley disc galaxies with the probability of a galaxy hosting a bar (hereafter \textit{bar probability}).

\subsection{Morphological catalog}

\citet{DominguezSanchez2018} catalog classifies morphologically $\sim$670,000 galaxies from SDSS galaxy images. The morphological classification procedure was based on using the Convolutional Neural Networks (CNNs) deep learning algorithms. This methodology is able to automatically learn the main characteristics from a training input data. To train the CNN models, \citet{DominguezSanchez2018} used the Galaxy Zoo 2 \citep[GZ2, ][]{Willett2013} and \citet{Nair2010} catalogs. GZ2 is a citizen science project that utilized non-professional volunteers to classify thousands of galaxies from SDSS galaxy images. The GZ2 final catalog contains $\sim 240,000$ galaxies with m$_r < 17$ and $z<0.25$ from SDSS DR7. The catalog from \citet{Nair2010} comprises $\sim14,000$ visually classified galaxies $-$ from SDSS DR4 within a redshift range of $0.01<z<0.1$. The performance of the CNN models were tested using two independent catalogs: the catalog from \citet{Huertas-Company2011}, which classifies $\sim 670,000$ SDSS galaxies by assigning to each galaxy a probability of belonging to one of four morphological classes (E, S0, Sab and Scd); and \citet{Cheng2011} catalog, which contains a combination of visual and automatic morphological classifications of 984 red passive galaxies. We refer the reader to \citet{DominguezSanchez2018} for more details about their morphological catalog.  

For each galaxy, \citet{DominguezSanchez2018} assign 9 different feature probabilities: disc probability (P$_{\text{disc}}$), edge-on probability (P$_{\text{edge-on}}$), bar signature probability through GZ2 training sample (P$_{\text{bar-GZ2}}$), bar signature probability through N10 training sample (P$_{\text{bar-N10}}$), merger probability (P$_{\text{merg}}$), bulge prominence probability (P$_{\text{bulge}}$), cigar-shaped probability (P$_{\text{cigar}}$), T-Type classification (connected with Hubble sequence, T-type) and S0 probability (P$_{\text{S0}}$). In this work we choose to characterize bar probability P$_{\text{bar-GZ2}}$ rather than P$_{\text{bar-N10}}$. This is because the Galaxy Zoo catalog provides many more galaxies to calibrate the CNN models and, therefore, we understand that P$_{\text{bar-GZ2}}$ is more reliable than P$_{\text{bar-N10}}$.

\subsection{Results}

\begin{figure}
\includegraphics[width=\columnwidth]{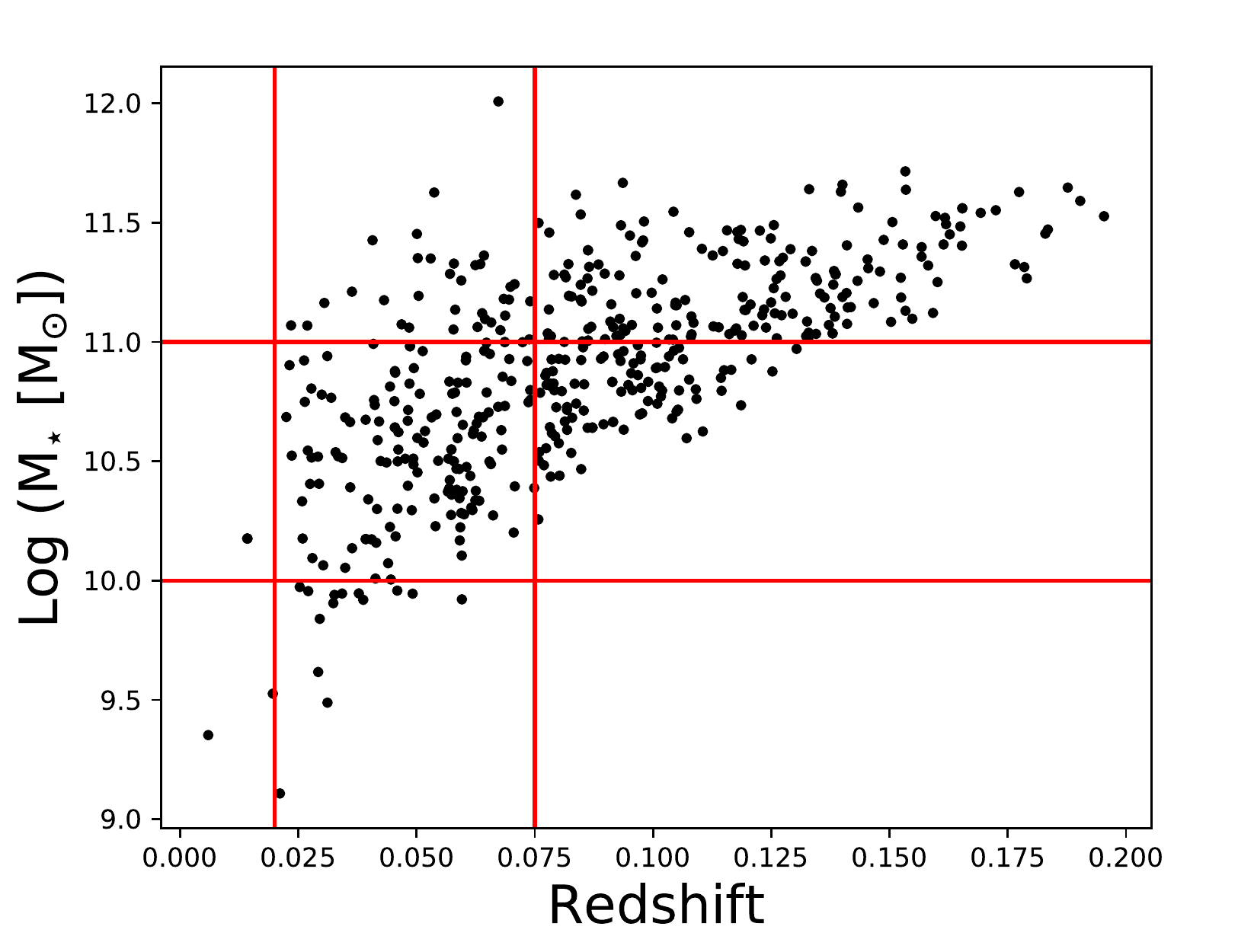}
\caption{Stellar mass distribution over redshift. The red lines indicate the completeness limits of our sample, with $10<\text{log}(\text{M}_{\star}[\text{M}_{\odot}])<11$ and $0.020<z<0.075$}.
\label{Stellar_mass_as_a_function_of_redshift}
\end{figure}

\begin{figure}
\includegraphics[width=\columnwidth]{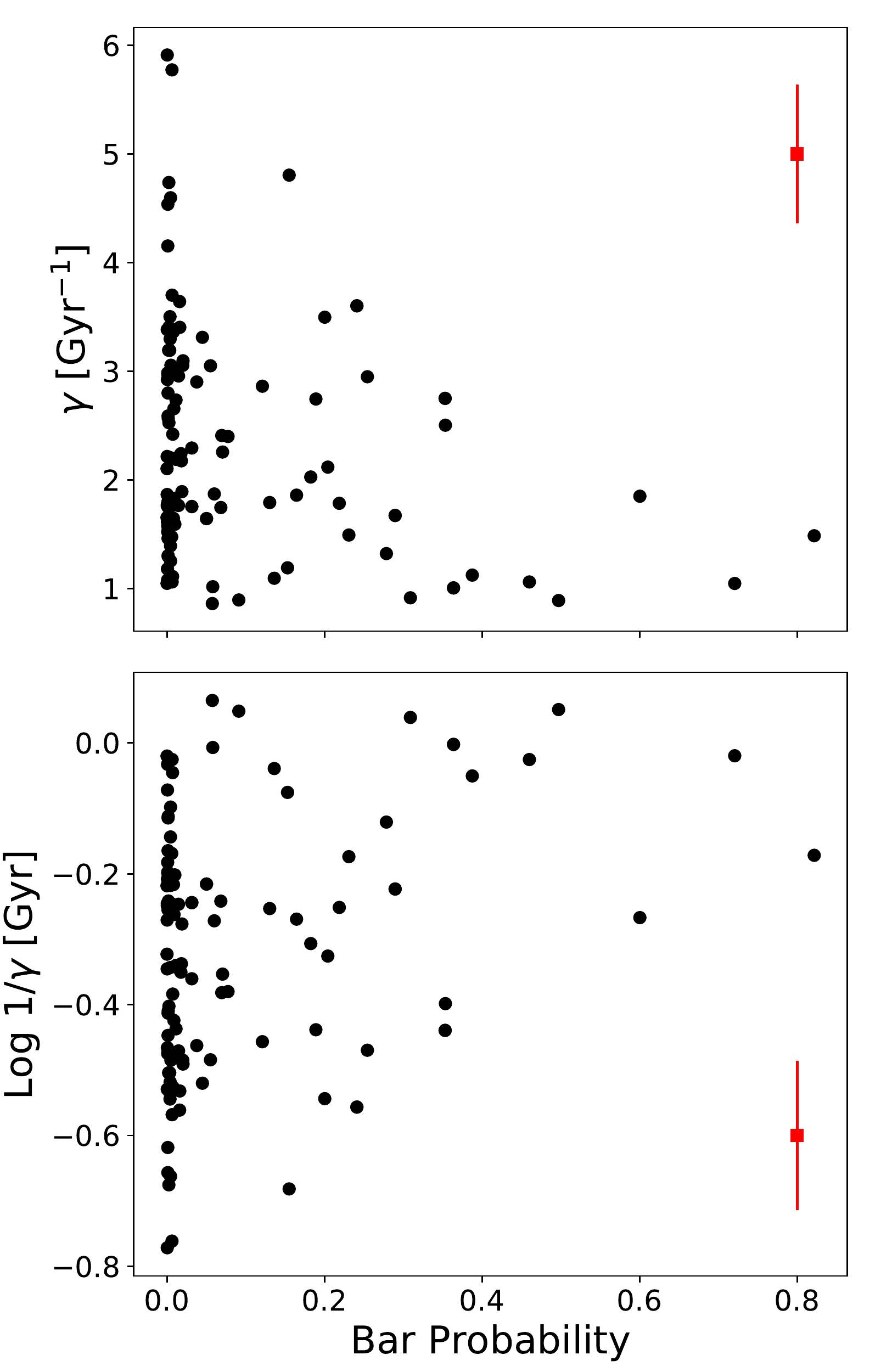}
\caption{Quenching indices ($\gamma$) and star formation quenching time-scales ($1/\gamma$) as functions of bar probability in our sample}. The red squares represent the average uncertainty in our sample.
\label{Gamma_as_a_function_of_p_bar_GZ2}
\end{figure}

From our cross-matched catalog we pick those with P$_{\text{merg}}<0.5$ and P$_{\text{disc}}>0.5$. With these conditions we aim at selecting predominantly disc galaxies that do not have a very close companion. We also constrain our sample in terms of stellar mass by choosing those within a mass range of $10^{10.0} < \text{M}_{\star} \ [\text{M}_{\odot}] < 10^{11.0}$. Our sample is complete at $z<0.075$, as shown in Fig. \ref{Stellar_mass_as_a_function_of_redshift}. \citet{Martin2017} discussed that the most massive green valley galaxies ($>10^{11}$~M$_{\odot}$) are probably red sequence galaxies that live at the centre of galaxy groups and clusters; these occasionally capture gas-richer satellites galaxies and consequently increase temporarily their star formation rate. As we are interested in galaxies that are only quenching their own star formation we chose to exclude these very massive green valley galaxies. Fig. \ref{Gamma_as_a_function_of_p_bar_GZ2} shows the star formation quenching time-scales of green valley galaxies as a function of bar probability. For low bar probability the quenching time-scales are spread from low to high values. However, green valley galaxies with higher bar probability (P$_{\text{bar}}>0.3$) tend to have lower quenching time-scales. We applied the Kolmogorov$-$Smirnov test to the distribution of quenching time-scales of green valley galaxies with P$_{\text{bar}}<0.1$ (very likely unbarred) and those of P$_{\text{bar}}>0.3$ (very likely barred) to check if these two sub-samples are significantly different. The resulting p-value of 0.006 strongly points against the null hypothesis, i.e., that they are drawn from the same population, showing that the distribution of quenching time-scales of green valley galaxies with low bar probability is indeed statistically different from those of green valley galaxies with high bar probability. We discuss the implications of our findings in the next Section.

\section{Discussions and conclusions}\label{discussions_and_conclusions}

Figure \ref{Quenching_timescales_from_JPLUS_photometry_and_SDSS_spectra_with_intrinsic_ext_corr} shows that our methodology with J-PLUS filters (F0395, F0410 and gSDSS) can recover quenching time-scales in green valley galaxies with a good precision in comparison with the methodology from  \citet{Martin2007}, which uses spectral indices from galaxy spectra. The quenching time-scales from our methodology converge with those from \citet{Martin2007} because the F0395, F04010 and gSDSS J-PLUS filters cover approximately the same region rest-frame wavelength used to define the D$_n$(4000) and H$_{\delta,A}$ spectral indices in galaxy spectra. This allows us to use the J-PLUS filters instead of galaxy spectra (Figs. \ref{F0395_g_as_a_function_of_Dn4000} and \ref{F0410_g_as_a_function_of_HdA}) to drive quenching time-scales.

Although both approaches are in agreement, Fig. \ref{Quenching_timescales_from_JPLUS_photometry_and_SDSS_spectra_with_intrinsic_ext_corr} shows a noticeable dispersion. Nevertheless, our method with J-PLUS filters recovers the quenching time-scales with a similar accuracy as that of \citet{Martin2007}, in spite of their usage of galaxy spectra. Therefore, Fig. \ref{Quenching_timescales_from_JPLUS_photometry_and_SDSS_spectra_with_intrinsic_ext_corr} demonstrates the potential of J-PLUS photometry to determine quenching time-scales for green valley galaxies in the local Universe. Furthermore, we expect much better constraints with the 56-narrow-band filter system of the upcoming J-PAS survey.

We measure quenching time-scales for a set of nearby green valley galaxies and correlate these time-scales with the probability of hosting a bar (Fig. \ref{Gamma_as_a_function_of_p_bar_GZ2}). The results indicate a tendency for green valley galaxies to have longer quenching time-scales for a higher bar probability. 

\begin{figure}
\includegraphics[width=\columnwidth]{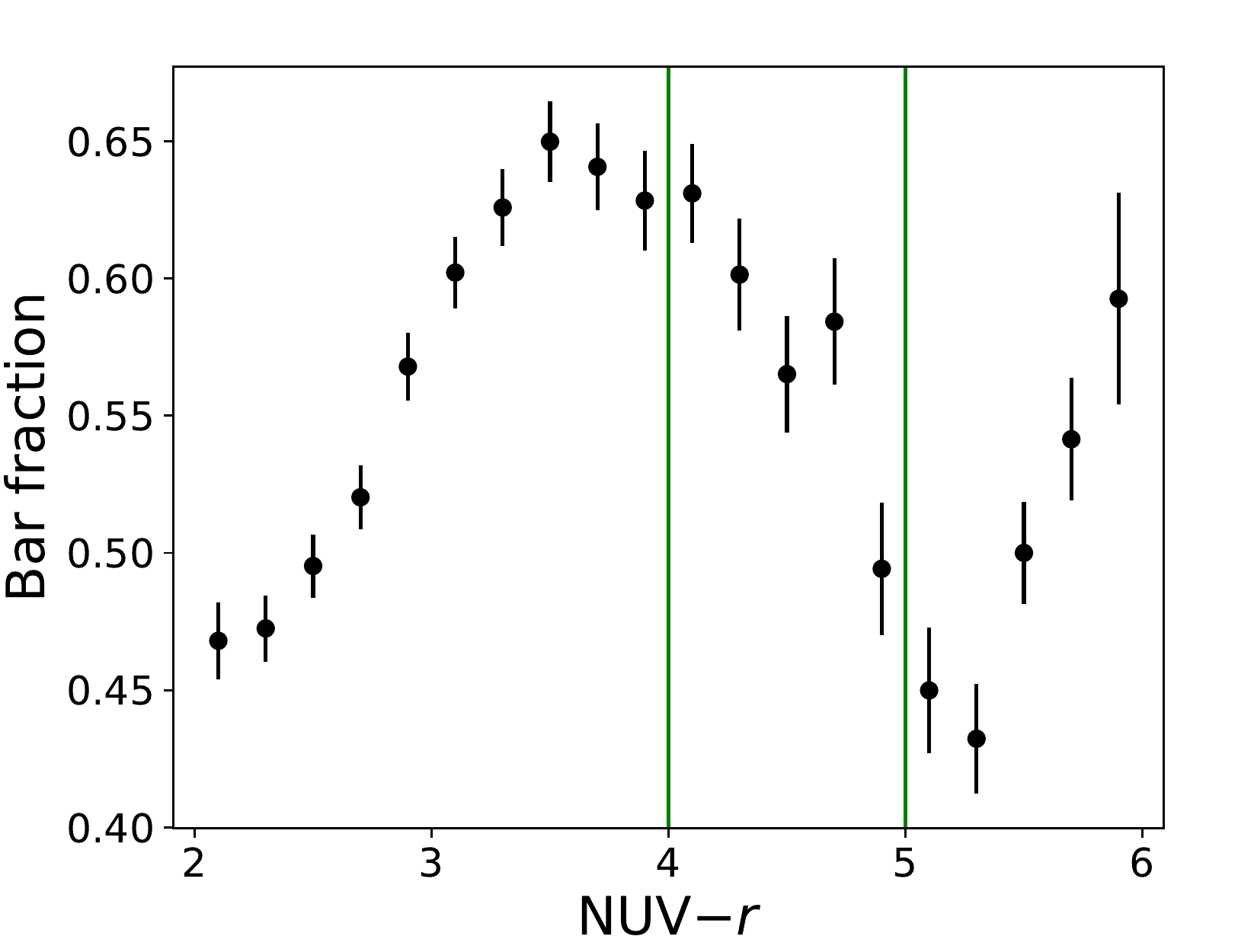}
\caption{Bar fraction as a function of dust corrected NUV$-r$ colour. We estimate the error bars as $\sqrt{f(1-f)/N}$ \citep[e.g., ][]{Sheth2008}, where $f$ is the bar fraction and $N$ is the total number of galaxies in each colour bin. The green vertical lines delimit the green valley region.}
\label{Bar_fraction_as_a_function_of_color}
\end{figure}

In order to better understand the impact of bars in quenching star formation in disc galaxies we also investigate the bar fraction evolution with NUV$-r$ colour (Fig. \ref{Bar_fraction_as_a_function_of_color}).  We find that the bar fraction behaviour seems to be have an interesting trend with galaxy colour. We note that both within the blue cloud and the red sequence the bar fraction increases towards redder colours by roughly 20\%. However, this increase is not continuous across the green valley. Within the green valley we note a significant drop in the bar fraction, resulting in a bar fraction dip at NUV$-r=6.0$. This general trend we observe in our sample is in agreement with previous works \citep[e.g., ][]{Cheung2013, Kelvin2018}.

\subsection{Bar formation and evolution}

N-body simulations show that bars are naturally formed within a few Gyrs or so in dynamically cold discs, although a dark-matter dominated center and gas-richness may delay this process somewhat \citep{Athanassoula1986, Athanassoula2002, Athanassoula2013, Aumer2017}. \citet{Sheth2008} found that bar fraction decreases with increasing redshift, consistent with galaxies having a tendency of being more gas rich at higher redshifts \citep{Genzel2010, Tacconi2010}. As galaxies become more dynamically-cold with time, the conditions become more favourable to the formation of a bar. This explains the high value of the local bar fraction in spiral galaxies \citep[$\sim$63\%; e.g., ][]{deVaucouleurs1963}.

We consider our results from Fig. \ref{Bar_fraction_as_a_function_of_color} in the light of this context. With gas richness dropping as we consider redder galaxies in the galaxy CMD, finding a lower bar fraction at the bluer end is reasonable. In fact, recent work by \citet{Kruk2018} shows that unbarred disc galaxies are bluer than barred ones. This is consistent with the increased bar fraction we observe in our sample for galaxies in the blue-cloud: galaxies with redder colours within the blue-cloud population appear to host bars more frequently. The increase in bar fraction that we observe towards redder red sequence galaxies may also be an extension of this logic. The dip in the bar fraction that appears to characterize the green valley represents an interruption to the otherwise global trend of increasing bar fraction with redder colours. The decrease in value of the bar fraction within the green valley suggests that processes that lead to the destruction of bars are at work.

Our results displayed in Fig. \ref{Bar_fraction_as_a_function_of_color} suggest that bars can form, be destroyed and form again as galaxies cross the colour- magnitude diagram. These same bar-destroying processes must either also be able to quench star formation in galaxies or act in parallel to the processes responsible for quenching star formation. It is interesting to determine the processes responsible for the destruction and the resurgence of bars in disc galaxies.

Over the last couple of decades different studies have pointed to the potential recurrence of the bar phenomenon, where a bar may form, grow in strength, get destroyed (even self-destroy) and ultimately reform in a galaxy disc \citep{Bournaud2002, Berentzen2004, Gadotti2006}. Hydrodynamical simulations have shown that the bar-induced increase in mass concentration in the central parts of barred galaxies can lead to bar self-destruction \citep{Roberts1979, Norman1996, Sellwood1999, Athanassoula2005}. It is worth noting that the amount of central mass concentrations required to effectively accomplish bar destruction are very large compared to observational constraints \citep{Debattista2000, Shen2004, Debattista2006}. Another way of destroying a bar is via tidal interactions with other galaxies \citep{Lee2012}, which may lead to the rapid quenching of star formation \citep{Nogueira-Cavalcante2018}. As galaxies become passive and gas-poor, the discs become susceptible again to develop a new stellar bar.

It is interesting to consider that the properties of bars in blue cloud barred galaxies have been shown to be different from those in the red sequence: bars in red discs tend to be longer and stronger whereas bars in star-forming discs are usually weaker and show an exponential light profile \citep{Elmegreen1985, Elmegreen1989}. This suggests that bars that persist through the galaxy transition from blue to red and through the green valley grow in size and strength. This suggests that, although a portion of the barred galaxies in the blue cloud goes through bar-destroying processes, another portion of the galaxy population has bars that persist through the green valley and continue to grow longer and stronger.

The combination of bar survival in some blue-cloud galaxies, together with the sequence of bar destruction and renewal in others would contribute to the increase in bar fraction within the red sequence galaxy population.

\begin{figure}
\includegraphics[width=\columnwidth]{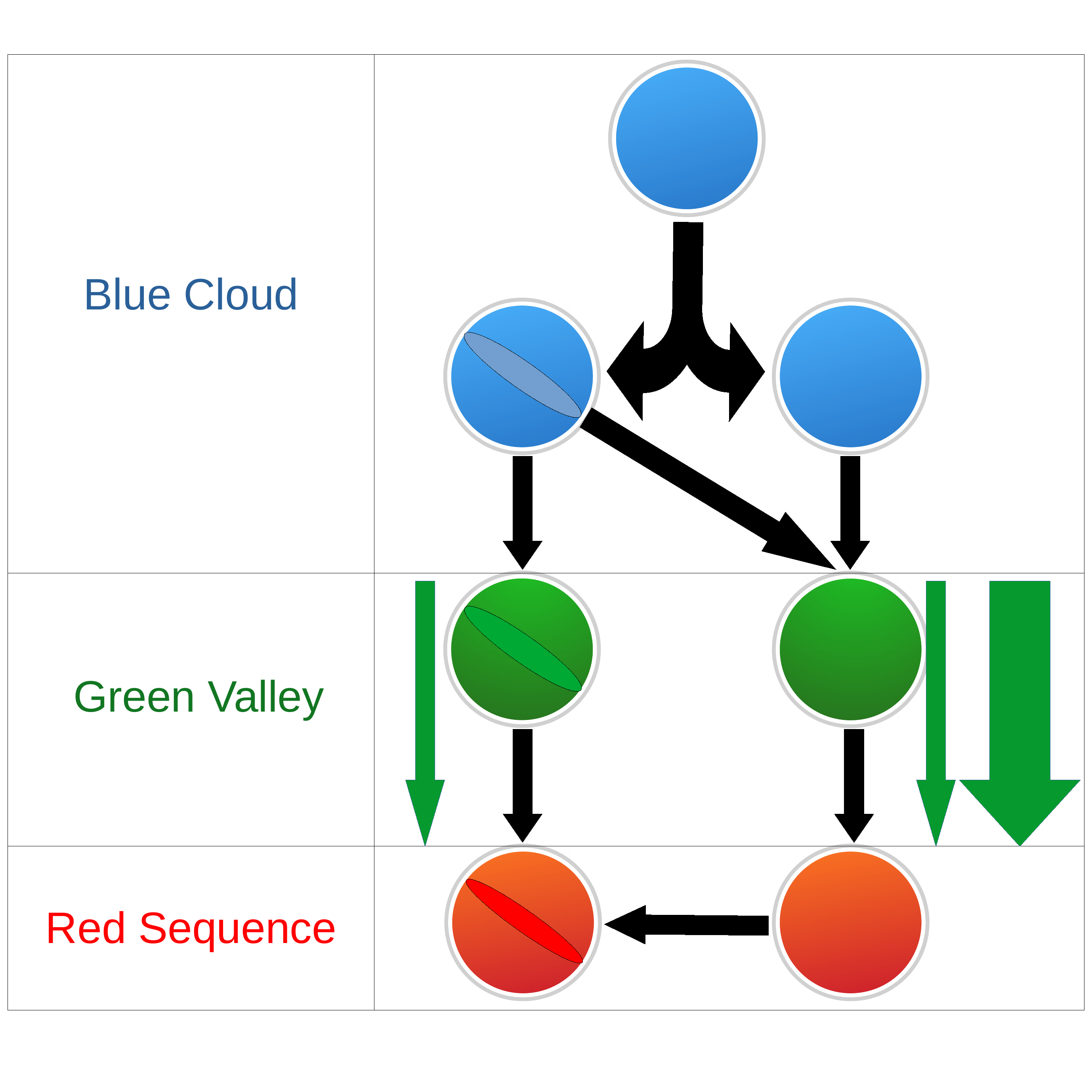}
\caption{Diagram showing a possibly scenario for barred and unbarred discs across the galaxy CMD. Galaxies start their evolution as star-forming blue unbarred galaxies and eventually may or may not develop a bar, depending on internal conditions. When these galaxies enter into the green valley phase, unbarred galaxies remain unbarred whereas many bars in barred galaxies are destroyed (e.g., via interactions). Unbarred disc galaxies can cross slowly or rapidly along the green valley (represented by narrow and wide green arrows, respectively) while barred galaxies only experience slow star formation quenching. When disc galaxies reach the red sequence, barred ones remain barred whereas unbarred ones can develop a new bar.}
\label{Diagram_barred_galaxies_evolution}
\end{figure}

We suggest a possible scenario for the transition across the green valley for galaxies with intermediate mass discs in the local Universe (see schematic representation in Fig. \ref{Diagram_barred_galaxies_evolution}). During the blue cloud phase a star forming galaxy may or may not develop a bar. When quenching processes start to act, barred galaxies may or may not lose their bars. As these galaxies transition through the green valley, the same star formation quenching processes (or other processes acting in parallel) may destroy their bars, driving the bar fraction down (Fig. \ref{Bar_fraction_as_a_function_of_color}). Galaxies whose bars do not persist through the star formation quenching phase, as well as galaxies that were unbarred within the blue cloud, they tend to transit through the green valley (and reach the red sequence) faster than barred green valley galaxies (Fig. \ref{Gamma_as_a_function_of_p_bar_GZ2}). Finally, after barred and unbarred galaxies reach the red sequence, barred galaxies tend to retain their bars whereas in unbarred galaxies the conditions become favourable for the formation/resurgence of bars.

\subsection{Bars and star formation quenching}

Many previous works have associated bars with quenching star formation. \citet{Schawinski2014} discussed that, among disc galaxies, stellar bars can accelerate star formation quenching. \citet{Masters2010} associated quenching with bars in order to explain the high bar fraction among passive disc galaxies. \citet{Gavazzi2015} noted that, for nearby disc galaxies, the bar fraction increases with stellar mass. As passive galaxies are more common among massive galaxies, these authors suggested that bars may play an important role in quenching star formation. \citet{Haywood2016} claimed that, in the case of our Galaxy, the stellar bar can stabilize the disc, preventing new star formation. Simulations also suggest that bars may efficiently quench star formation, decreasing by a factor of 1~Gyr the star formation rate time-scale \citep{Khoperskov2018}. 

Although these works suggest that bars induce a faster star formation quenching, our results displayed in Fig. \ref{Gamma_as_a_function_of_p_bar_GZ2} show that green valley disc galaxies with high bar probability tend to quench star formation  slowly, which in agreement with \citet{Coelho2011}, who found that the stellar population in bulges of barred galaxies shows an excess of populations younger than $\sim4$ Gyr, when compared to bulges in unbarred galaxies, for the same galaxy masses. Fig. \ref{Gamma_as_a_function_of_p_bar_GZ2} also shows that green valley disc galaxies with low bar probability can also quench star formation very slowly and we find that many disc galaxies with low bar probability have quenching time-scales greater than those in green valley galaxies with high bar probability. On the other hand, our results suggest that in galaxies with high bar probability the processes which quickly quench star formation are either absent or sustained by the gravitational potential of the host galaxy, which is sufficient to preserve the bar and, simultaneously, the star formation against these processes. \citet{Nogueira-Cavalcante2018} found a similar result for green valley barred galaxies at intermediate redshifts ($0.5<z<1.0$), with barred galaxies, displaying the longest star formation quenching time-scales.

The increase of the bar fraction in the red sequence region (NUV$-r>5$ in Fig. \ref{Bar_fraction_as_a_function_of_color}) reflects the greater ease that the red discs have to form stellar bars. Therefore, bars are probably not associated with accelerated star formation quenching, contrary to simulations, which suggest that bars can increase gas consumption in isolated discs in comparison with unbarred ones \citep{Khoperskov2018}. However, real galaxies are complex structures where several quenching mechanisms (where the majority of them can be bar-destroyers) happen simultaneously. Inevitably, if a disc galaxy remains isolated long enough and keeps the disc dynamically cool, a bar will be formed. Therefore, the barred systems can be interpreted as those where violent physical processes, that quench star formation rapidly, are absent.

\subsection{Bar colours impact on the quenching time-scales estimations}

In this work we determine the $\gamma$ parameter through the NUV$-r$, F0395$-$gSDSS and F0410$-$gSDSS colours. A reasonable explanation for the systematic low $\gamma$ values (high quenching time-scales) for green valley barred galaxies is that F0395-gSDSS and F0410-gSDSS colours in these galaxies are redder for a specific NUV-$r$ value, in comparison with green valley unbarred galaxies. These reddening F0395-gSDSS and F0410-gSDSS colours could be caused either by dust extinction or by older stellar population in the bar region in barred galaxies.

\begin{figure}
\includegraphics[width=\columnwidth]{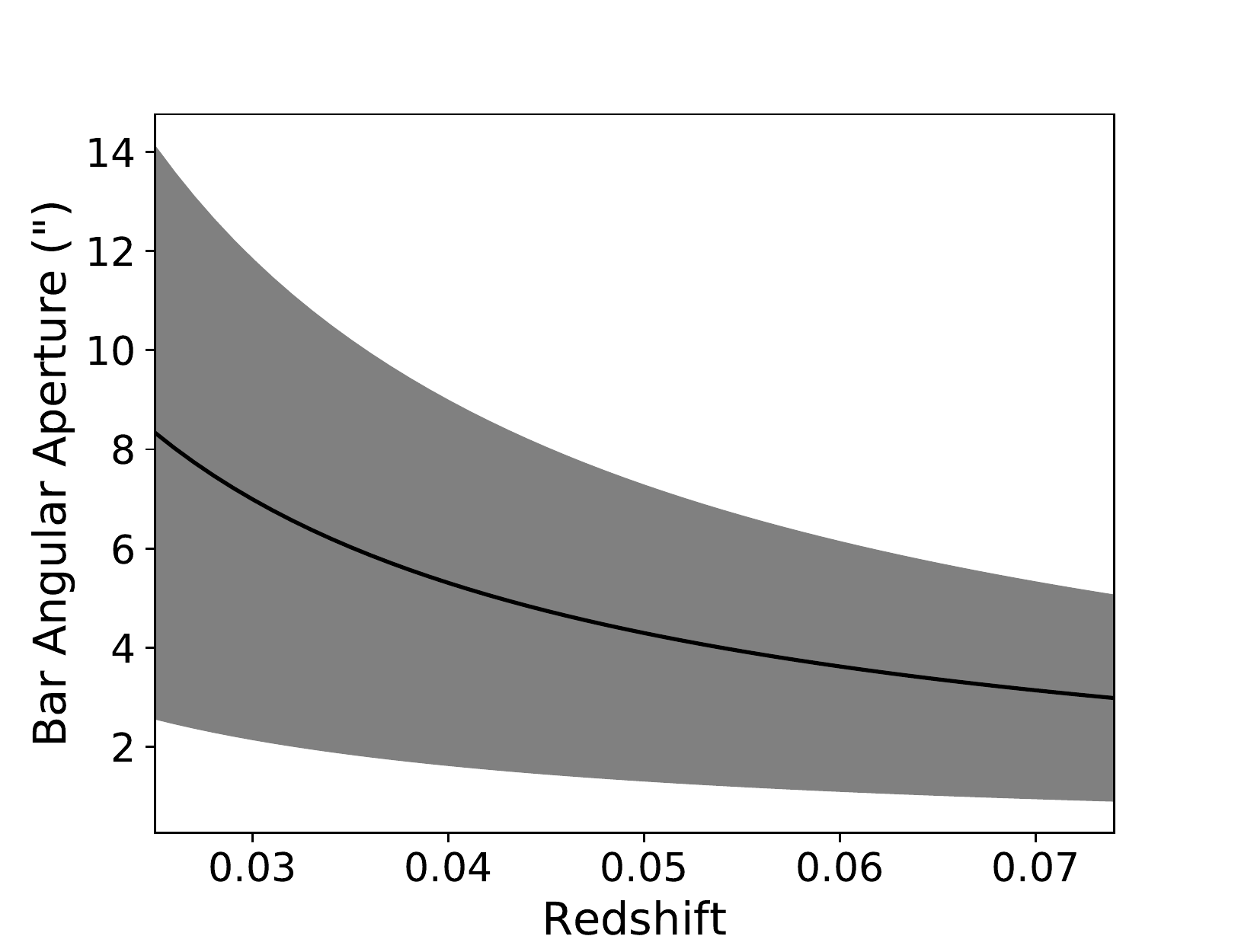}
\caption{Bar angular aperture of typical barred galaxies as a function of redshift. The solid black line and the gray region represent the average value and the standard deviation of the bar sizes of barred galaxies in the local Universe \citep{Menendez-Delmestre2007}.}
\label{angular_aperture_as_a_function_of_redshift_for_bars}
\end{figure}

To test the dust extinction hypothesis, we analyze the F0395-gSDSS and F0410-gSDSS colours in different angular apertures, in order to compare the central regions (bar region) of unbarred and barred green valley galaxies. The size of bars in a typical disc galaxy in the local Universe is $4.2 \pm 2.9$ kpcs \citep[e.g., ][]{Menendez-Delmestre2007}. Fig. \ref{angular_aperture_as_a_function_of_redshift_for_bars} shows the necessary angular aperture as a function of redshift to cover the bar region of a typical barred galaxy in the local Universe. The aperture photometries avaiable in J-PLUS DR1 are 0.8", 1.0", 1.2", 2.0", 3.0", 4.0" e 6.0". From Fig. \ref{angular_aperture_as_a_function_of_redshift_for_bars} we choose the 3.0", 4.0" e 6.0" angular apertures, which cover almost entirely the bar region of typical local barred galaxies over the redshift range of our galaxy sample.

\begin{figure*}
\includegraphics[width=\textwidth]{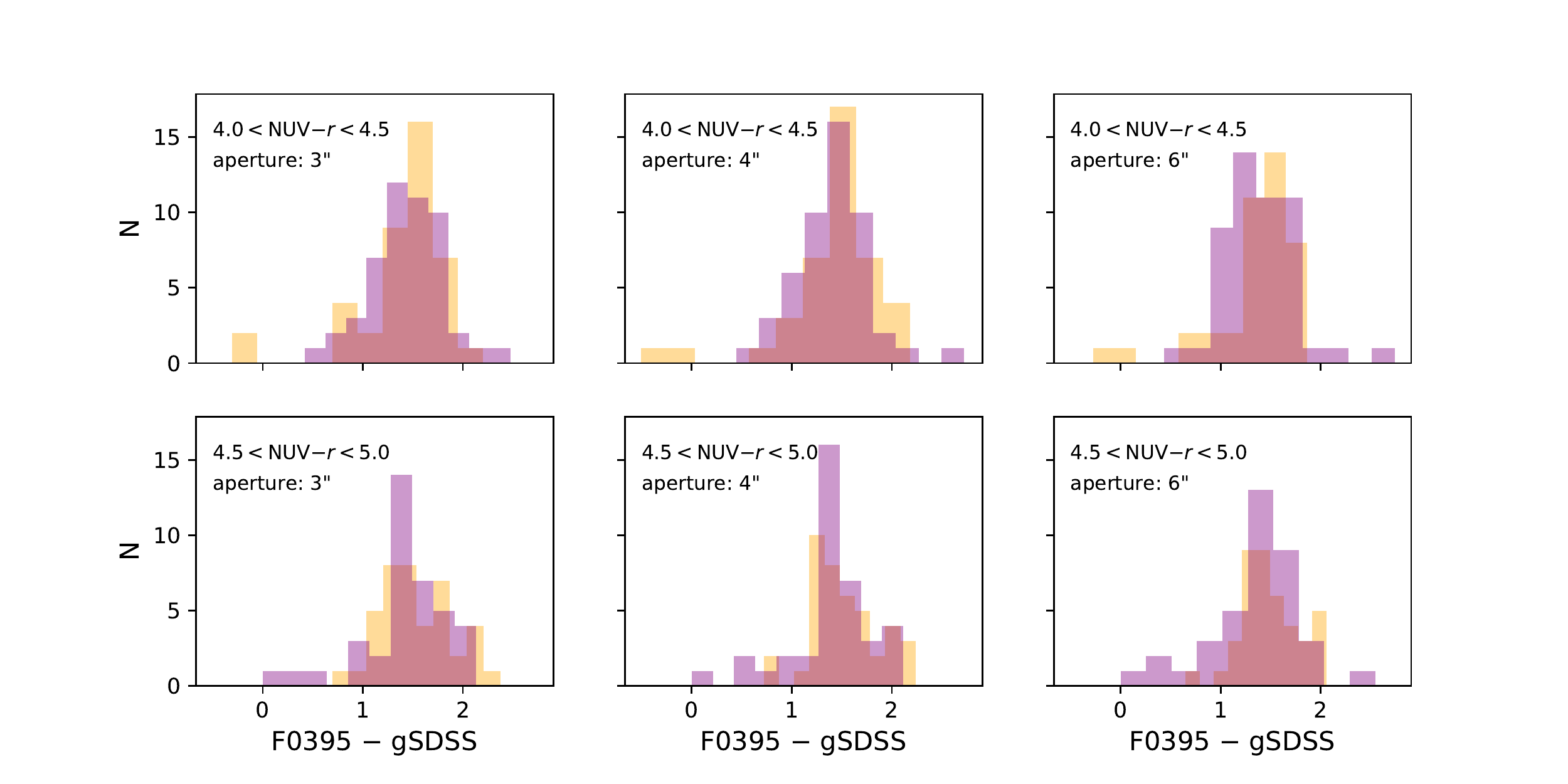}
\caption{F0395$-$gSDSS colour distributions for barred (purple) and unbarred (orange) green valley galaxies for different angular apertures and NUV$-r$ colour ranges.}
\label{histograms_F0395_gSDSS}
\end{figure*}

\begin{figure*}
\includegraphics[width=\textwidth]{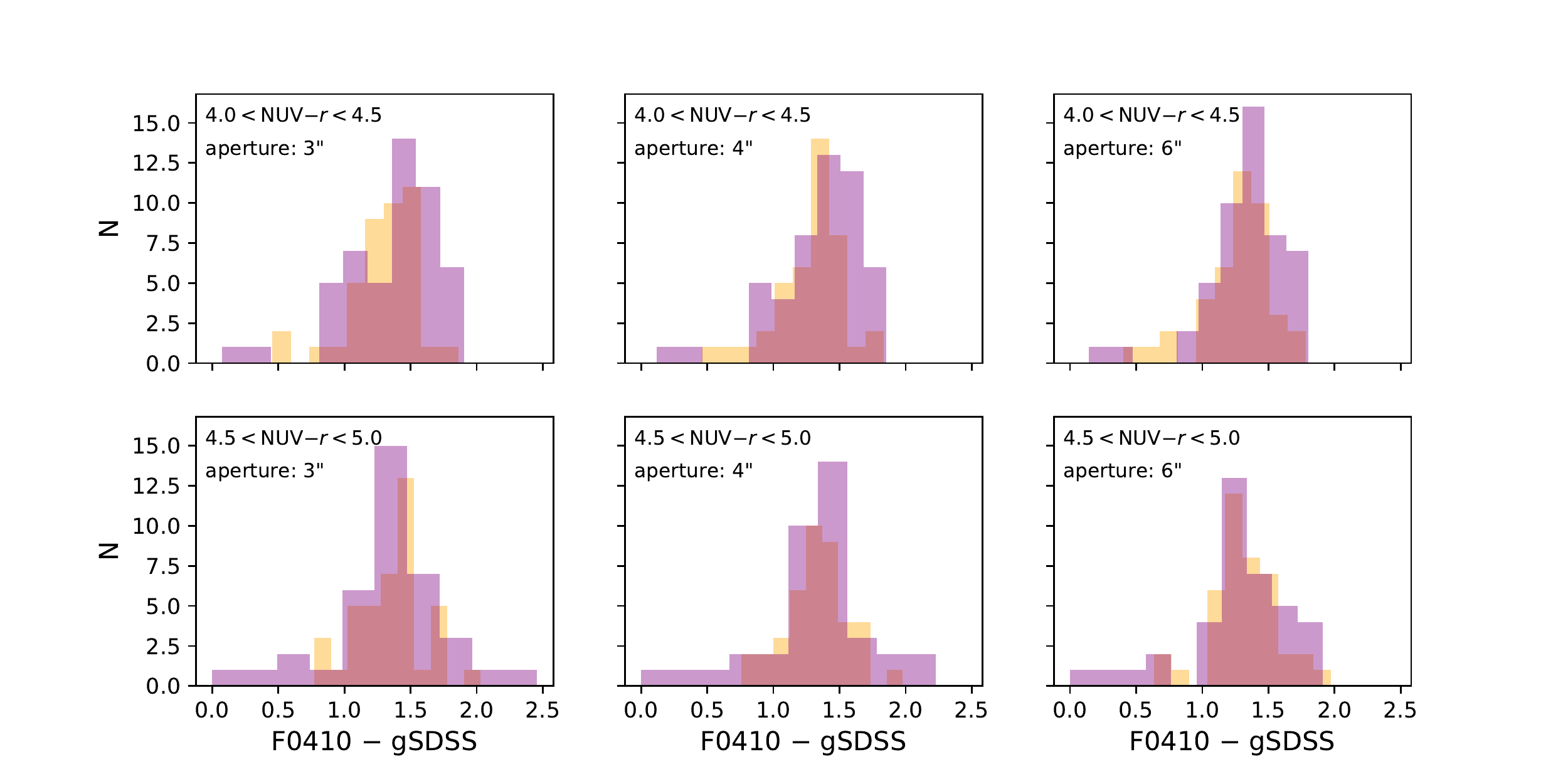}
\caption{Same in the Fig. \ref{histograms_F0395_gSDSS} for the F0410$-$gSDSS colours.}
\label{histograms_F0410_gSDSS}
\end{figure*}

Figures \ref{histograms_F0395_gSDSS} and \ref{histograms_F0410_gSDSS} show the distribution of F0395$-$gSDSS e F0410$-$gSDSS colours, respectively, for barred and unbarred green valley galaxies of our sample. From these Figs. we do not find colour excess for the green valley barred galaxies in the regions delimited by the angular apertures. Therefore, Figs. \ref{histograms_F0395_gSDSS} and \ref{histograms_F0410_gSDSS} suggest that the low $\gamma$ values (Fig. \ref{Gamma_as_a_function_of_p_bar_GZ2}) for green valley barred galaxies are due a more a more continuous star formation, that simultaneously affects the NUV$-r$, F0395$-$gSDSS e F0410$-$gSDSS colours, and not due dust extinction in the bar region.

To test for a possible bias that could appear from a tentative older stellar population in the bar region we compare qualitatively the luminosity-weighted age of stellar populations from the bar region and other parts of the galaxy using MaNGA data of a sample of face-on barred green valley galaxies. The MaNGA survey \citep{Bundy2015, Yan2016} is a SDSS-IV \citep{Blanton2017} project that aims to obtain spatially resolved spectroscopy for $\sim$10000 local galaxies ($z < 0.15$), with a spectral resolution of $R\sim2000$, from 4000$\AA$ to 8500$\AA$, using the BOSS spectrograph \citep{Smee2013, Drory2015} on the Sloan Telescope \citep{Gunn2006}. MaNGA galaxies come from the NASA Sloan Atlas catalogue \citep[NSA, ][]{Blanton2005}, which are representative of all the galaxy populations (blue cloud, green valley and red sequence).

We obtain luminosity-weighted ages from the public MaNGA \texttt{FIREFLY} catalog \citep{Goddard2017}. \texttt{FIREFLY} \citep[Fitting IteRativEly For Likelihood analYsis, ][]{Wilkinson2015} is a spectral fitting code which can obtain several stellar population properties from galaxy spectra, such as stellar ages and metallicities. For the spectral fitting \texttt{FIREFLY} assumes: foreground
Milky Way reddening dust maps by \citet{Schlegel1998} and extinction curve by  \citet{Fitzpatrick1999}; intrinsic reddening by dust in each galaxy by \citet{Wilkinson2015}; and stellar population models by \citet{Maraston2011}, with MILES stellar library \citep{Sanchez-Blazquez2006} and Kroupa initial mass function \citep{Kroupa2001}.

\texttt{FIREFLY} code both estimates luminosity-weighted and mass-weighted stellar ages. The former is sensitive to recent star formation whereas the latter is affected by the cumulative galaxy evolution. We decided to use luminosity-weighted age because recent star formation changes directly galaxy colours and therefore can influence the estimations of quenching time-scales.

\begin{figure*}

\includegraphics[width=6.cm]{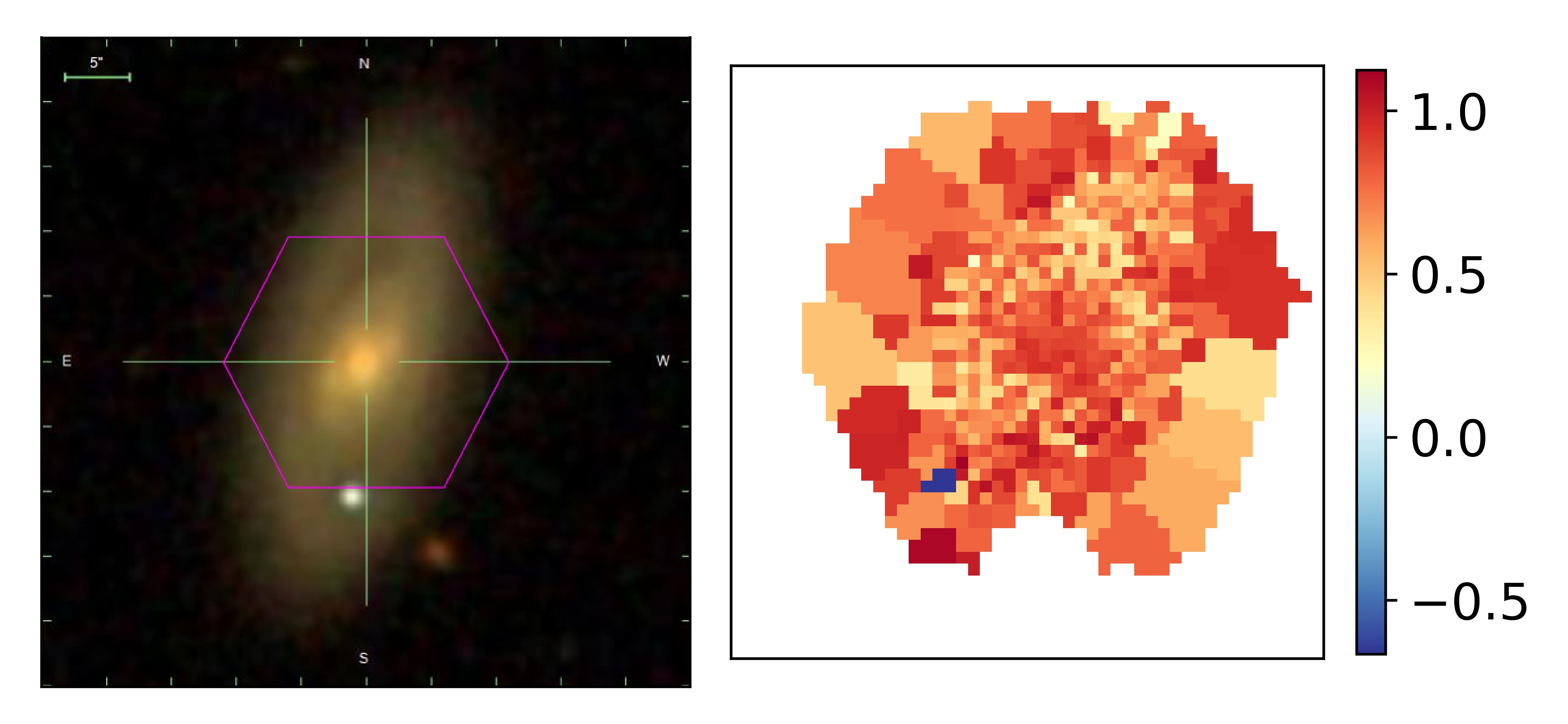}
\includegraphics[width=6.cm]{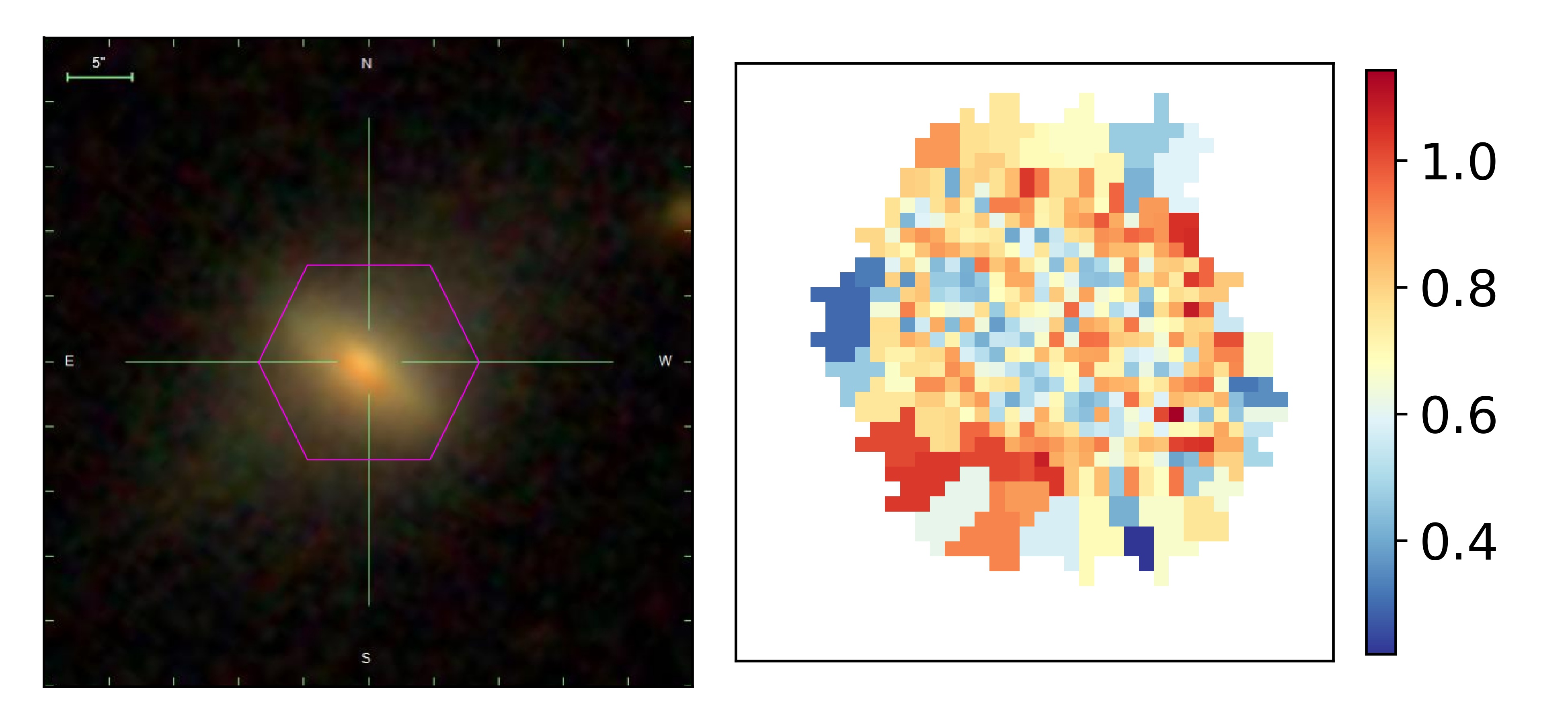}
\includegraphics[width=6.cm]{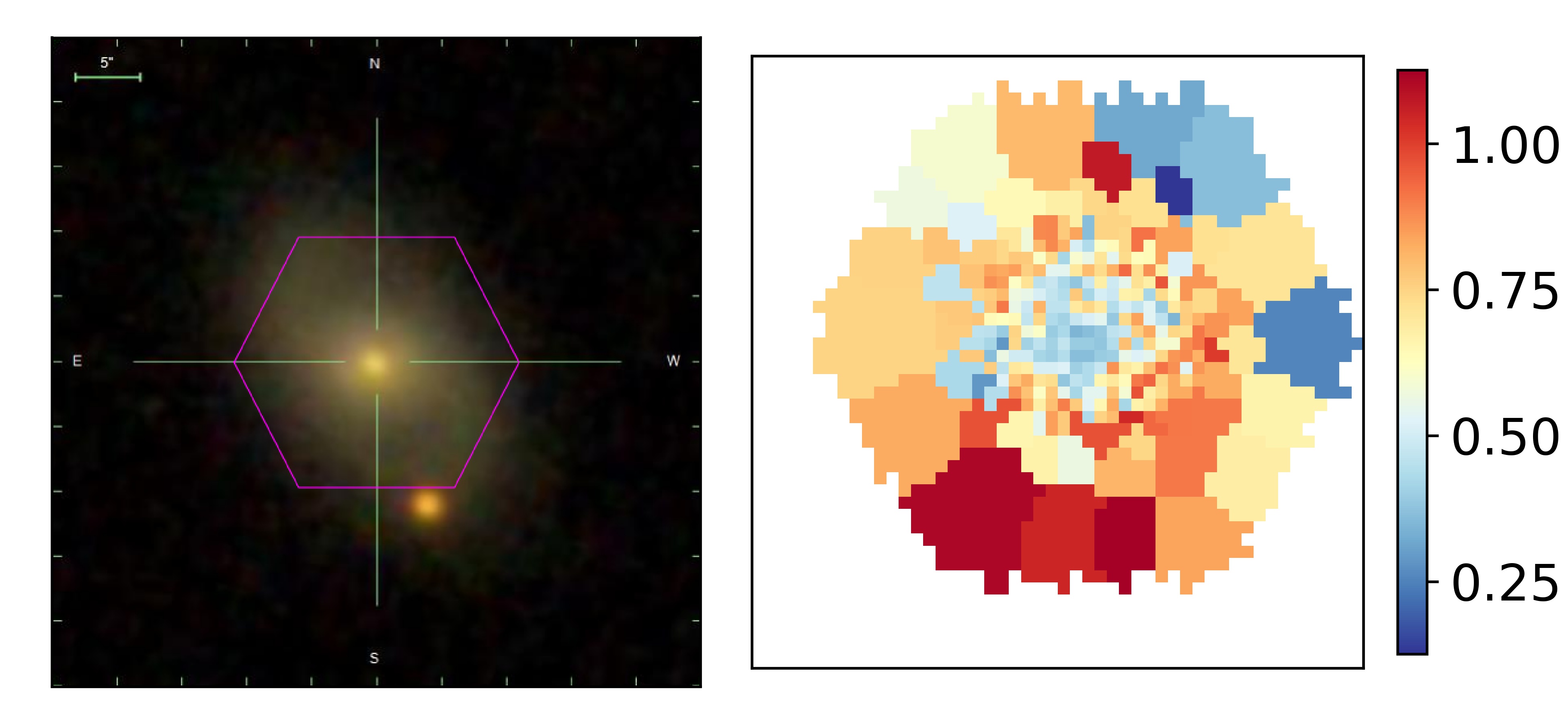}
\includegraphics[width=6.cm]{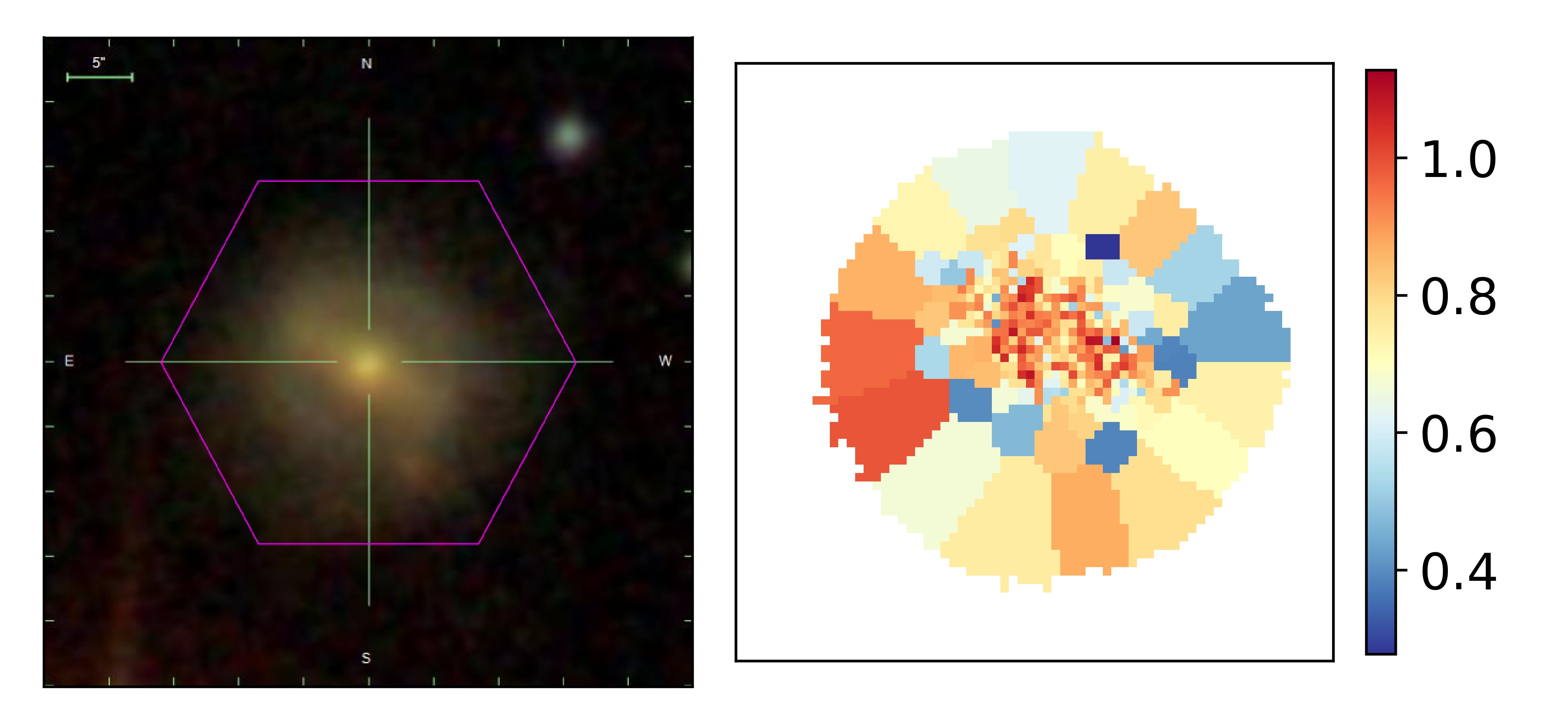}
\includegraphics[width=6.cm]{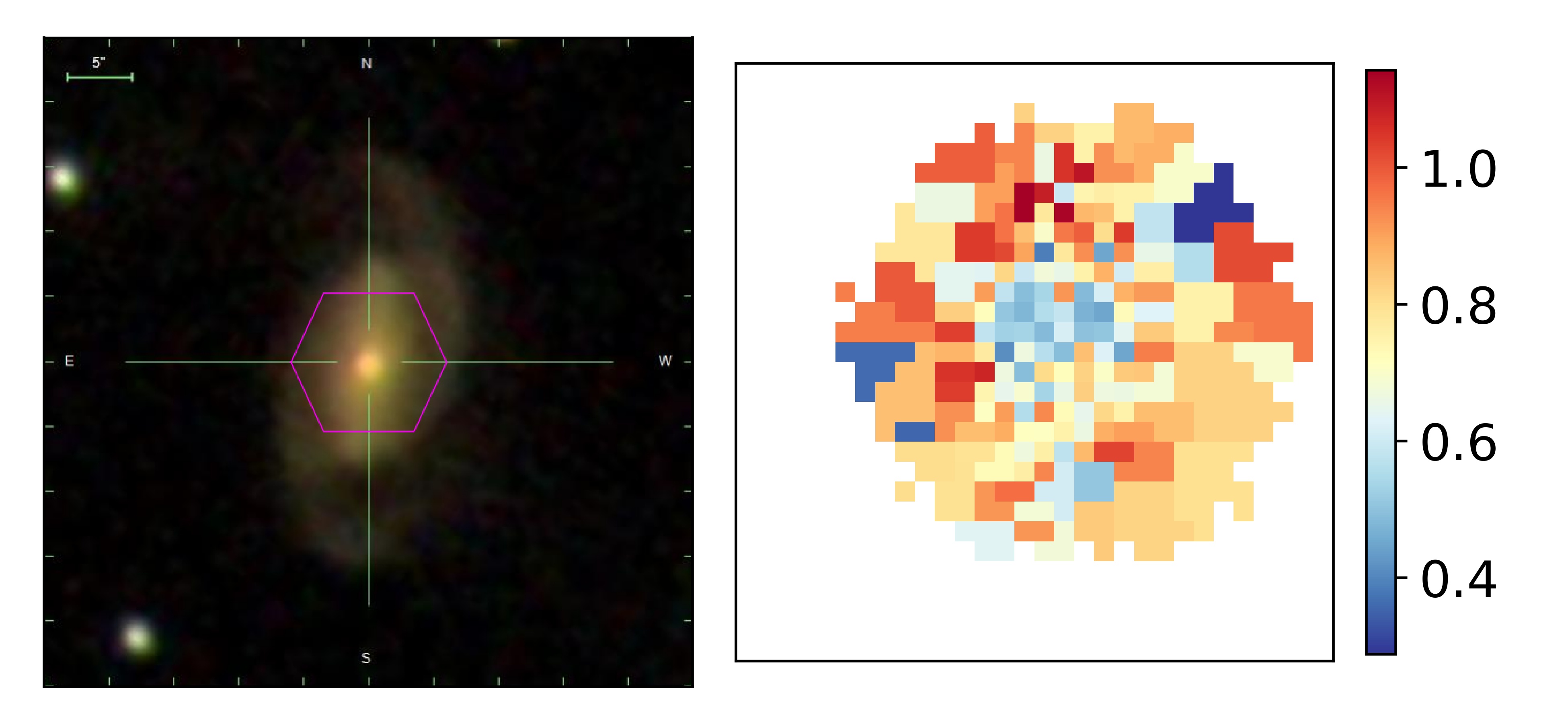}
\includegraphics[width=6.cm]{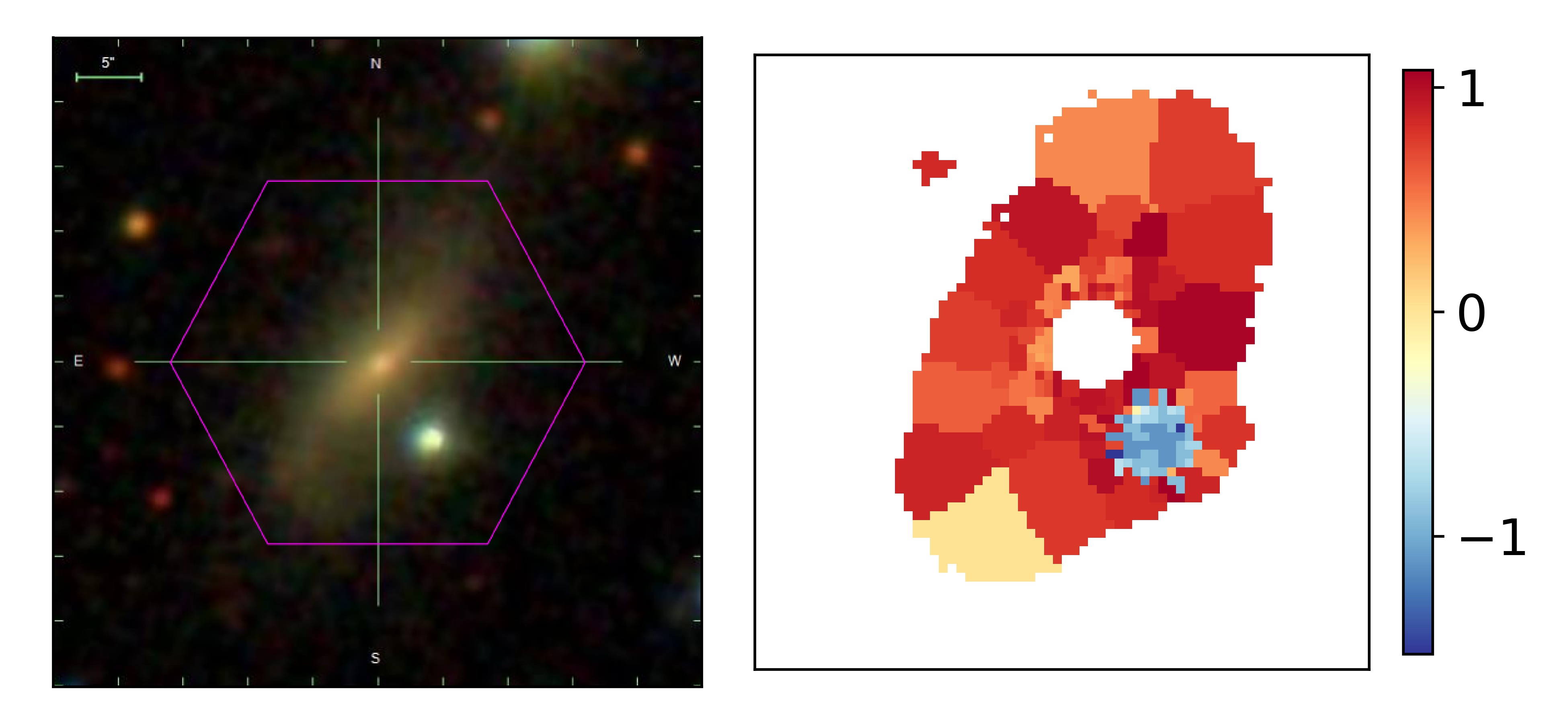}
\includegraphics[width=6.cm]{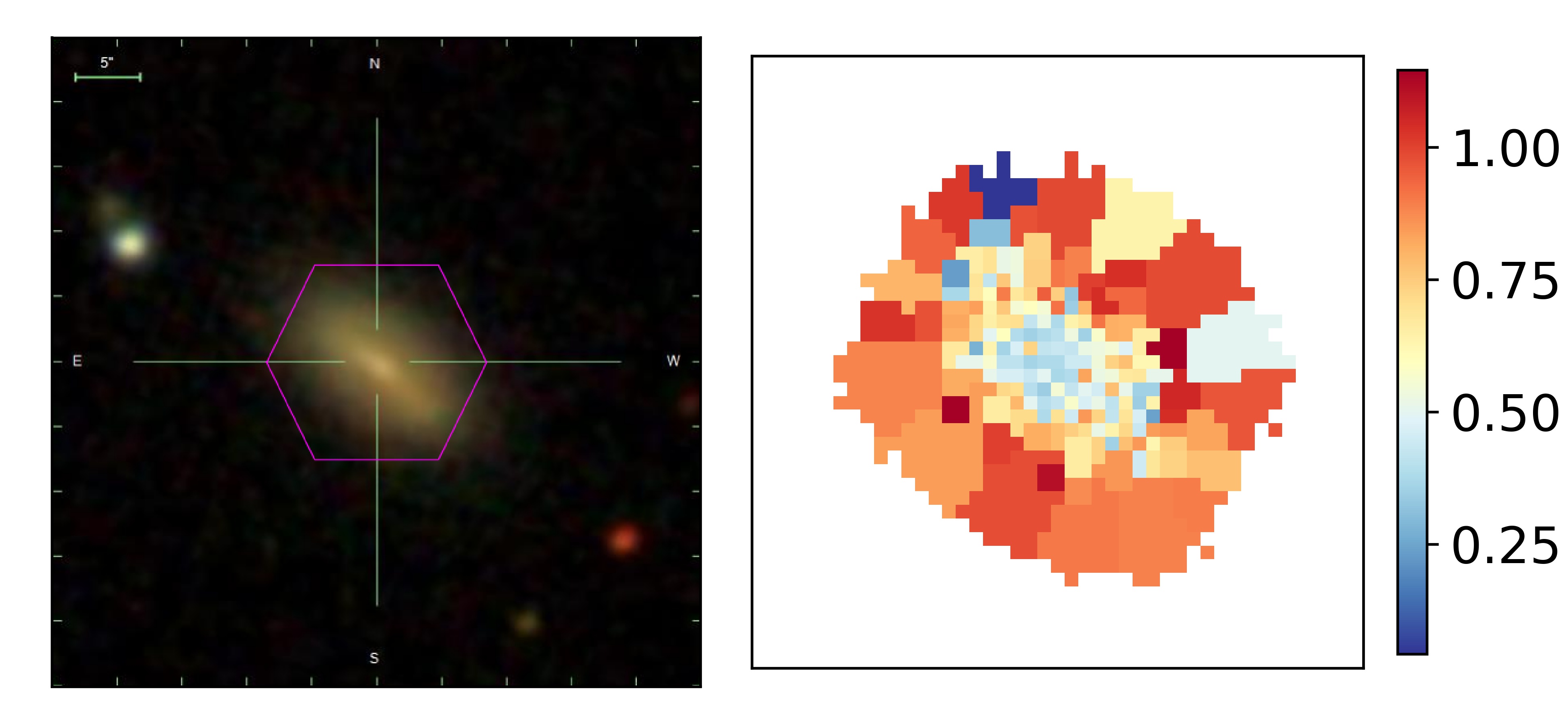}
\includegraphics[width=6.cm]{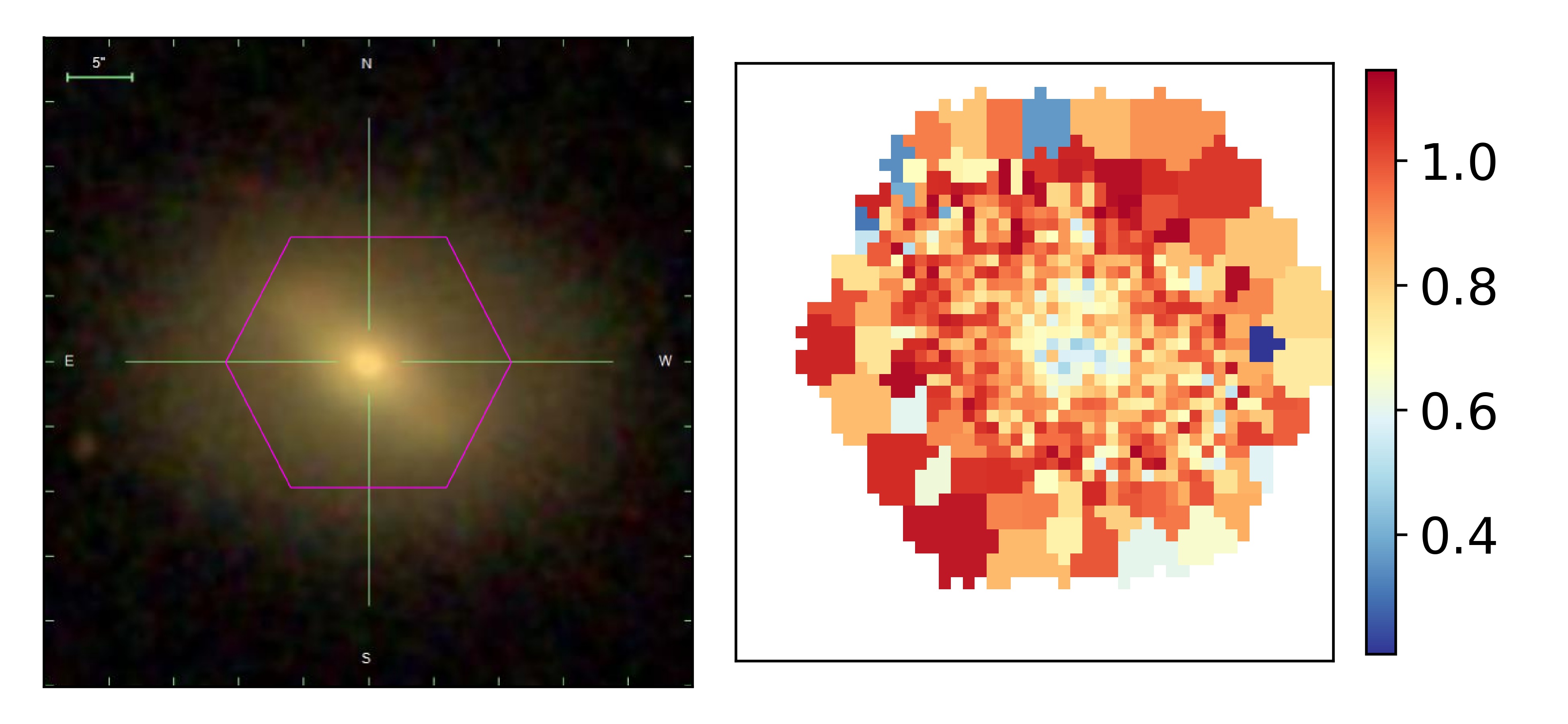}
\includegraphics[width=6.cm]{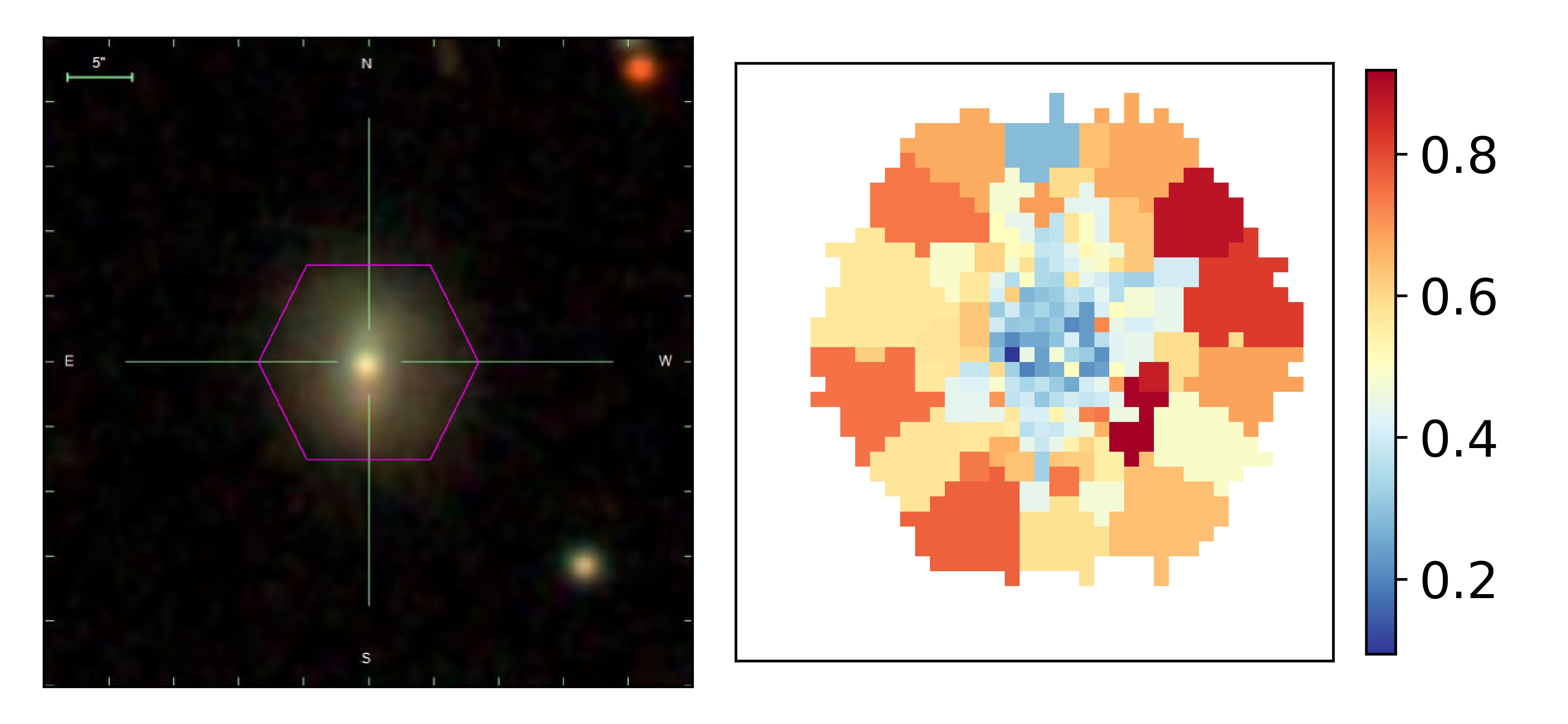}
\includegraphics[width=6.cm]{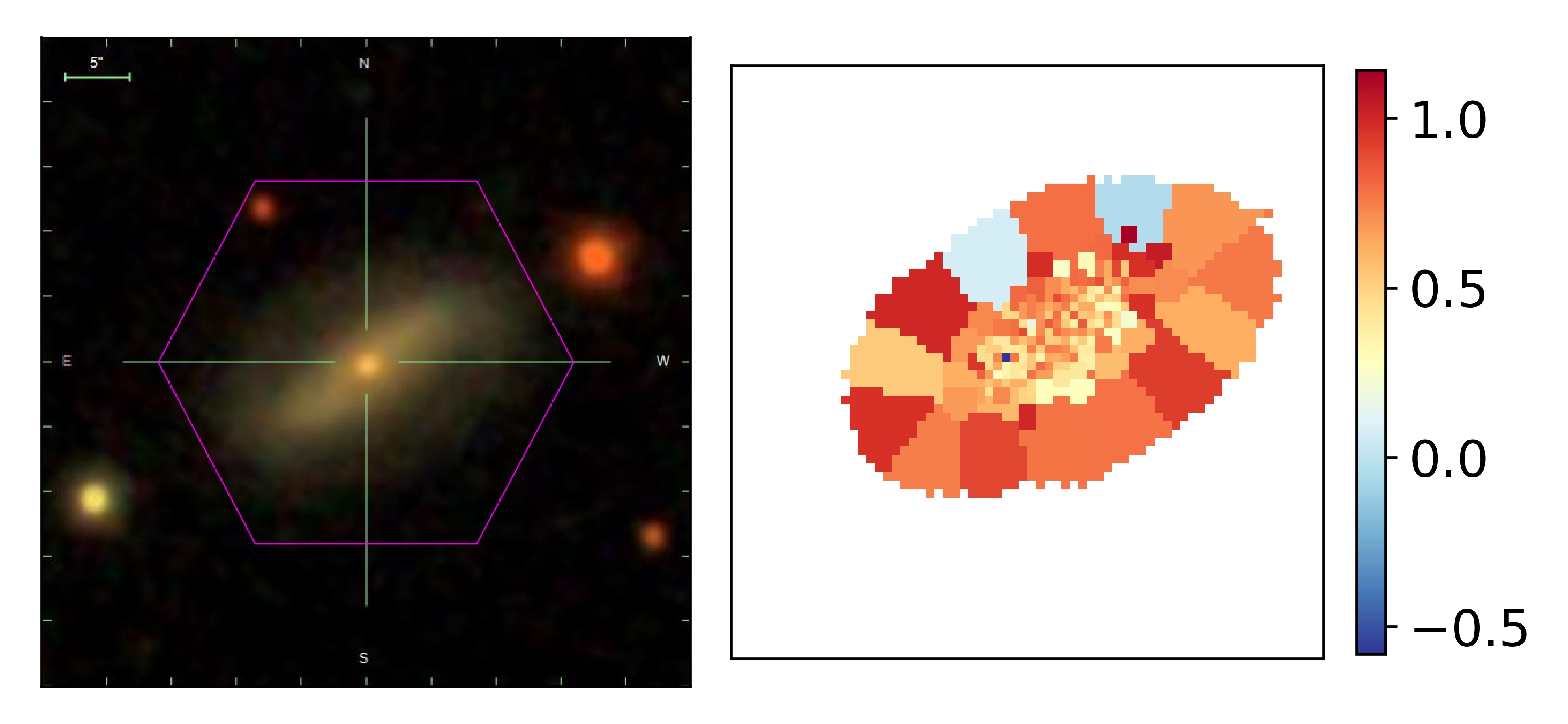}
\includegraphics[width=6.cm]{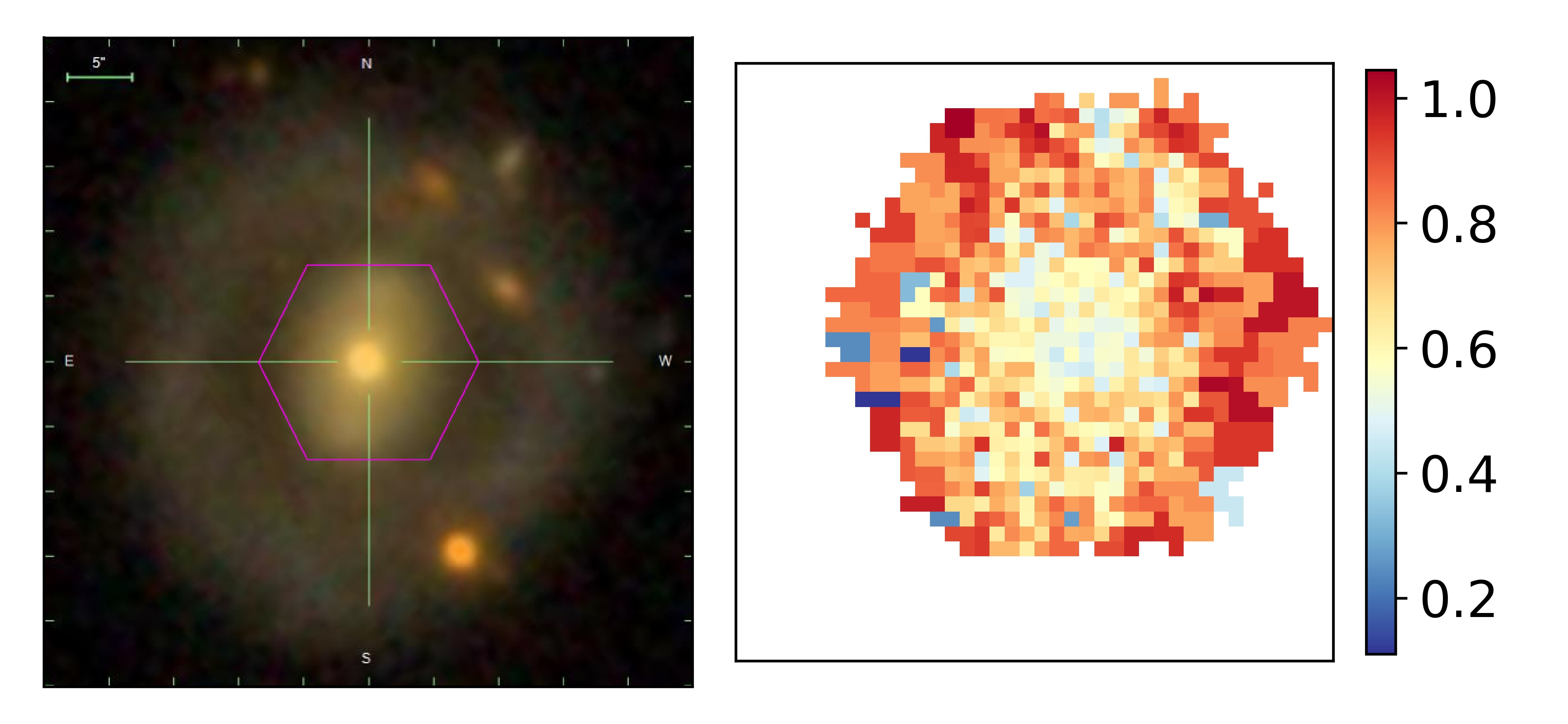}
\includegraphics[width=6.cm]{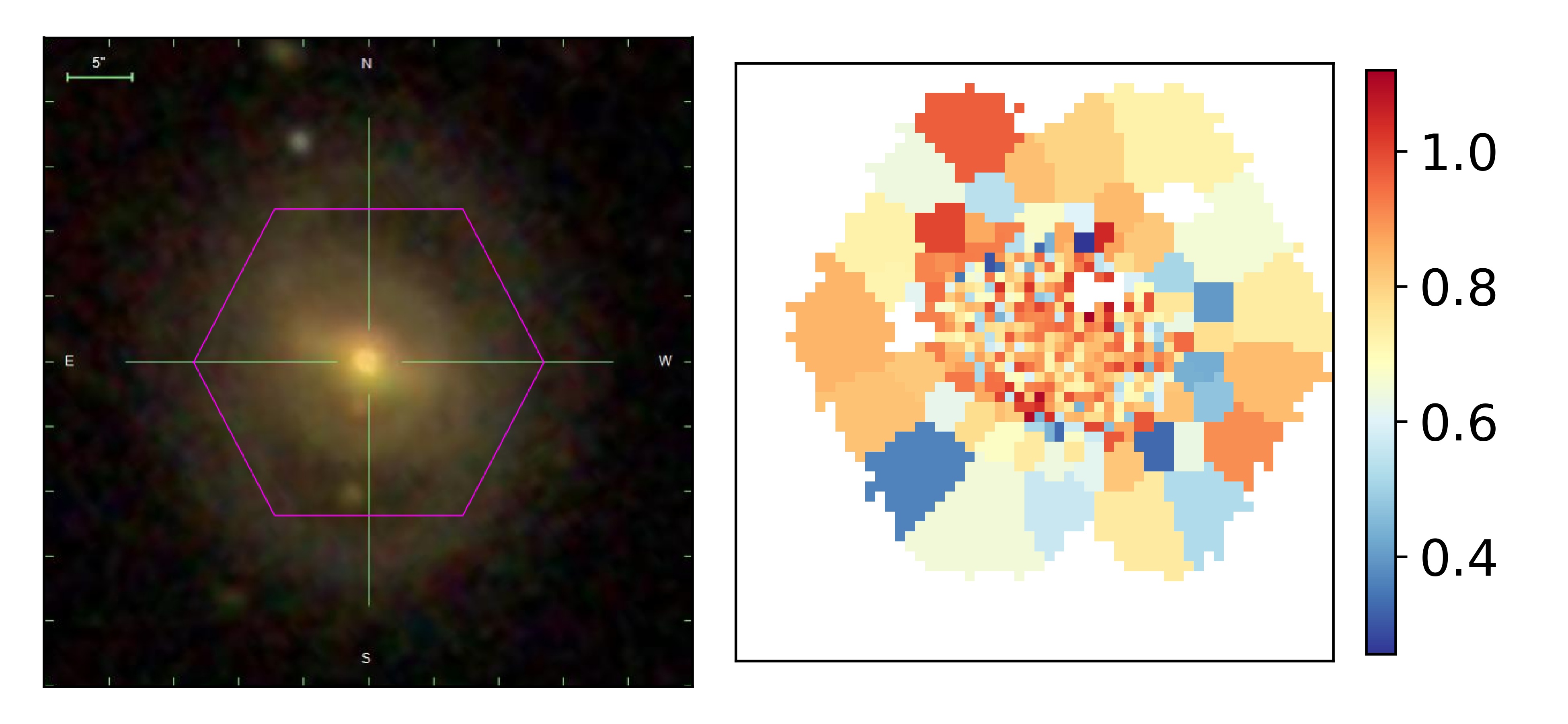}
\includegraphics[width=6.cm]{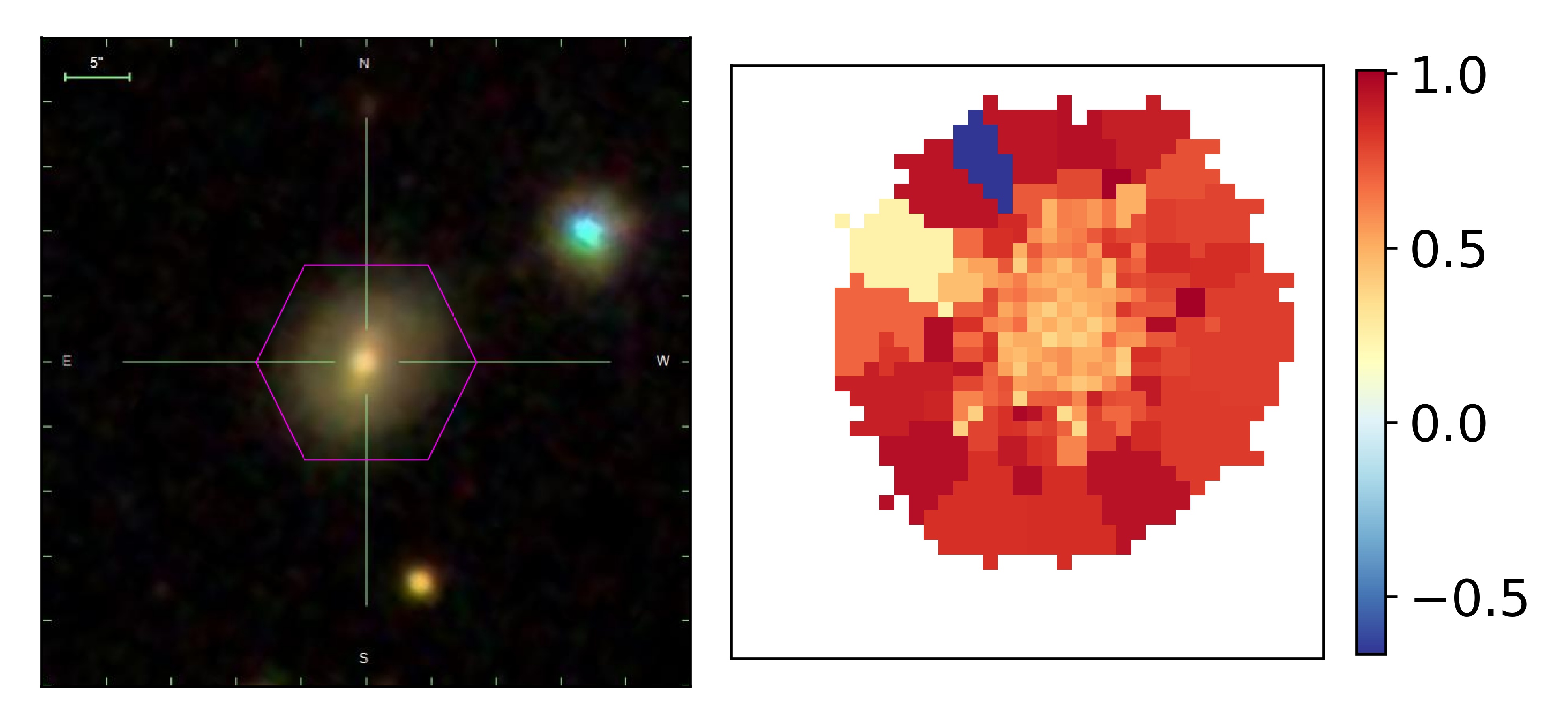}
\includegraphics[width=6.cm]{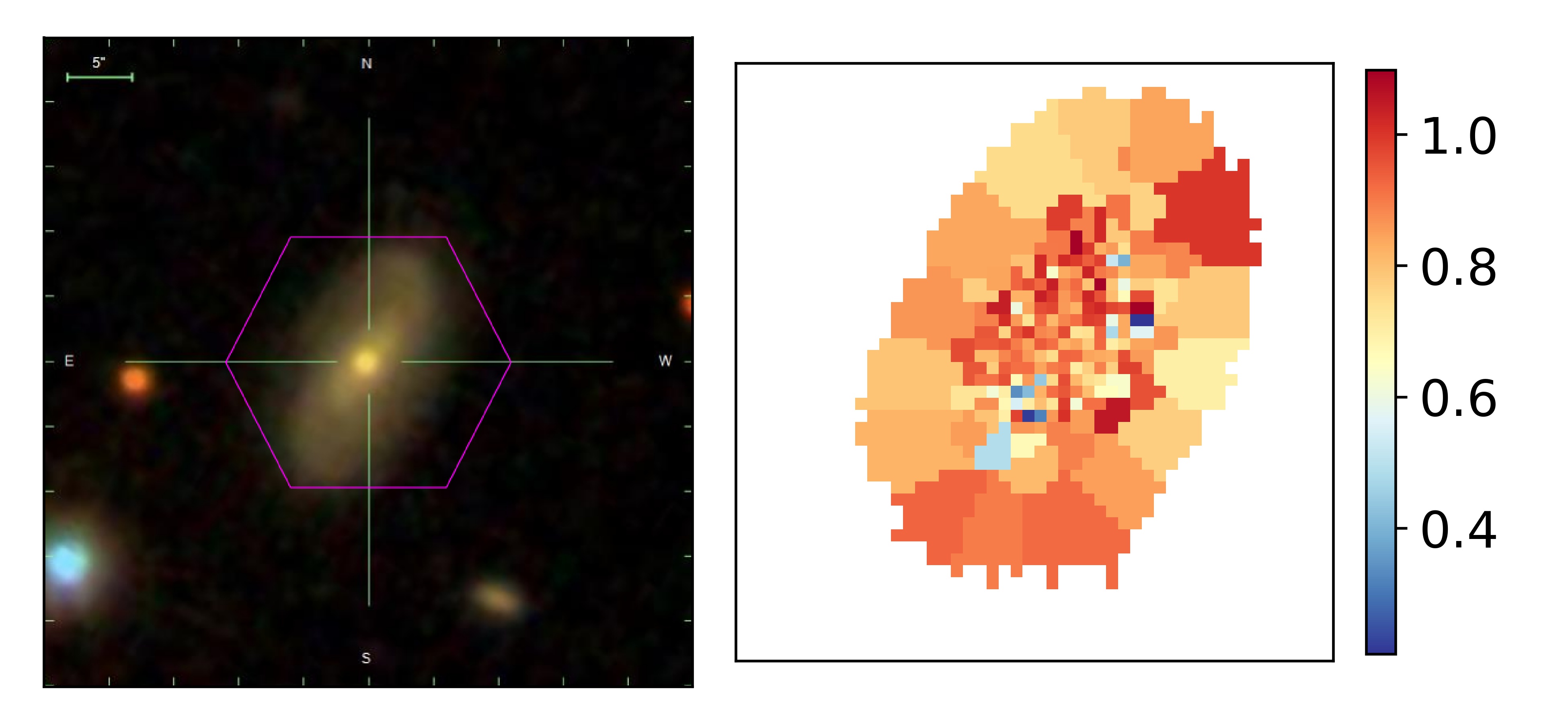}
\includegraphics[width=6.cm]{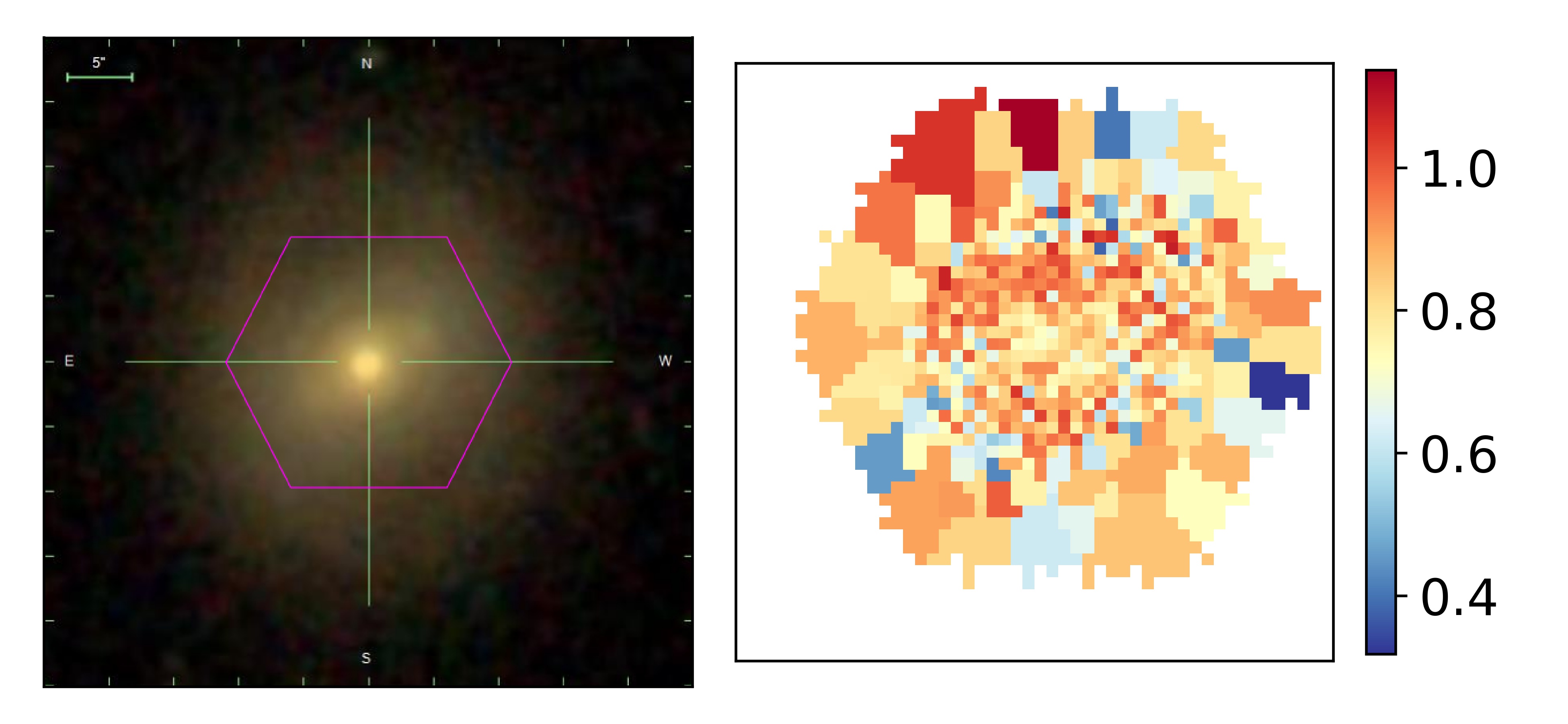}

\caption{SDSS galaxy images and luminosity-weighted stellar ages of a sample of face-on barred green valley galaxies. The purple hexagon indicates the MaNGA field-of-view and the luminosity-weighted ages is expresed in log$_{10}$(age[Gyr]).}
\label{galaxy_images_and_luminosity_weighted_ages_of_barred_green_valley_galaxies}
\end{figure*}

Fig. \ref{galaxy_images_and_luminosity_weighted_ages_of_barred_green_valley_galaxies} shows SDSS galaxy images and luminosity-weighted age maps of a sample of face-on barred (P$_{\text{bar}}>0.2$) green valley galaxies identified in the MaNGA \texttt{FIREFLY} catalog. From this figure we do not find any clear indication that the bar region is systematic older (redder) than the other parts of the galaxy. Therefore the quenching time-scales from Fig. \ref{Gamma_as_a_function_of_p_bar_GZ2} for barred green valley galaxies is very likely to be reflecting the overall galaxy evolution.

\subsection{Photo-z impact on the quenching time-scales estimations}

\begin{figure}
\includegraphics[width=\columnwidth]{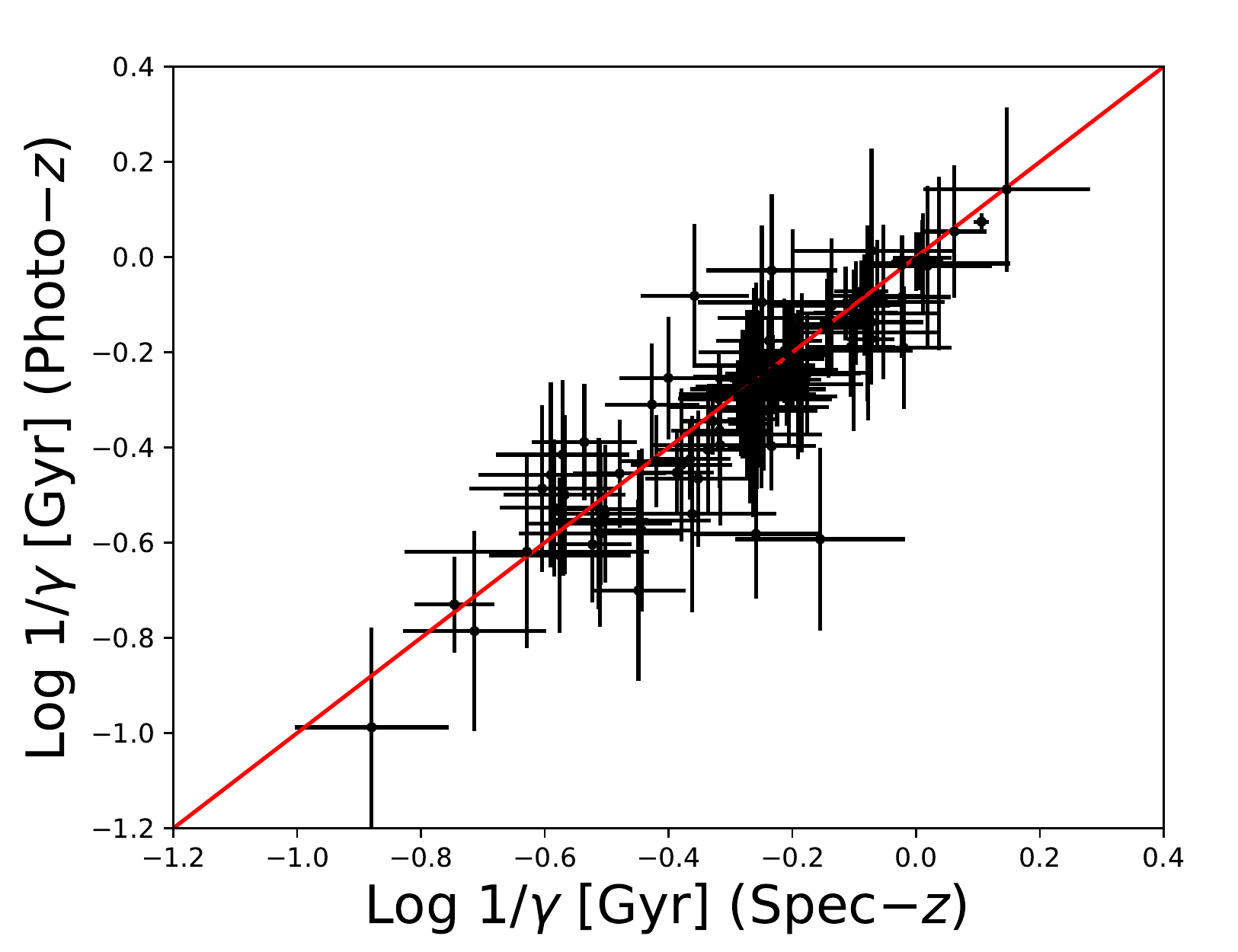}
\caption{Comparison of quenching time-scales based on photometric and spectroscopic redshifts.The spearman correlation and RMSE are, respectively, 0.89 and 0.09.}
\label{gamma_photoz_as_a_function_of_gamma_specz}
\end{figure}

Another legacy of J-PLUS will be marked by photometric redshift determination of galaxies. We explore the impact of photo$-z$ uncertainties when computing quenching time-scales with J-PLUS data. The photometric redshifts are calculated using the BPZ code \citep{Benitez2000}. Fig. \ref{gamma_photoz_as_a_function_of_gamma_specz} shows the correlation between the quenching time-scales obtained with spectroscopic redshifts with those obtained with photo$-z$. We find that for the majority of green valley galaxies we can recover approximately the same time-scales, demonstrating that the photo$-z$ uncertainties impact is very small in quenching time-scales estimations.  

\section{Summary}\label{summary}

The scarcity of galaxies between the blue and red peaks on galaxy CMD (green valley) and the growth of the red sequence (at least from $z\sim2$) indicate that galaxy transition across the green valley is rapid and continuous. The identification of the physical processes that are able to transform galaxies from star-forming to passive systems still remains a challenge to galaxy formation and evolution models. Stellar bars are one of the possible physical processes that could accelerate gas depletion and consequently quench star formation faster. In this work we adapted the method of \citet{Martin2007} $-$ which is able to estimate star formation quenching time-scales in green valley galaxies based on spectroscopic data $-$ to the photometric data from J-PLUS and GALEX surveys. Through this method, we measured quenching time-scales of nearby green valley galaxies ($0.022<z<0.075$) and  we correlated these time-scales with the probability of these galaxies hosting a bar (bar probability).

Our main results are:

\begin{enumerate}
    
    \item Our method using photometric data is able to estimate quenching time-scales at the same precision level as that from \citet{Martin2007}, which assesses quenching time-scales from galaxy spectra.
    
    \item We find a strong correlation between quenching time-scales and bar probability of green valley disc galaxies. Quenching time-scales among green valley galaxies with low bar probability have a large scatter whereas for high bar probability green valley disc galaxies tend to have long quenching time-scales. These results suggest that violent quenching processes, which are able to simultaneously destroy the bars and quench star formation quickly are absent or the gravitational potential of the host galaxy is strong enough to maintain both the stellar bar and a more stable star formation.  
    
    \item The apparent bar bimodality with NUV$-r$ colour indicates potential different bar phases. During the blue cloud phase bars are naturally formed, which is reflected by the increase of bar fraction with colour, until NUV$-r\sim3.5$. After this peak, bar fraction quickly decreases, reaching a minimum at the beginning of red sequence phase (NUV$-r\sim5.2$). This suggests that the same quenching processes that move galaxies across the green valley can also destroy the bars or bar-destroyers processes happen simultaneously with quenching processes. Red sequence discs can recover their bars, as indicated by the increase of bar fraction in NUV$-r>5.2$.  
    
\end{enumerate}

Our analysis of quenching time-scales as a function of bar probability in nearby green valley galaxies and bar fraction as a function of NUV$-r$ colour suggests an evolutionary scenario (Fig. \ref{Diagram_barred_galaxies_evolution}) where massive disc galaxies ($10^{10}< \text{M} \ [\text{M}_{\odot}]<10^{11.5}$) quench their star formation in slow mode wheres unbarred galaxies can quench their star formation both in slow and fast modes. Future surveys, such as the recently started J-PAS, maybe can address this scenario in galaxy-dense environments.

\begin{acknowledgements}
We thank the anonymous referee for contributing to improving our paper. JPNC was supported by a Institutional Capacity Program grant from CNPq $-$ the Brazilian National Council for Scientific and Technological Development $-$ within the Ministry of Education of Brazil and Ministry of Science, Technology, Innovations and Communication. RD acknowledges support from the CNPq through BP grant 312307/2015-2, and the Financiadora de Estudos e Projetos - FINEP grants REF. 1217/13 -01.13.0279.00 and REF 0859/10 - 01.10.0663.00 for hardware support for the J-PLUS project through the National Observatory of Brazil and CBPF. PC acknowledges support from FAPESP 2018/05392-8, CNPq 305066/2015-3 and USP/Cofecub 18.1.214.1.8/40449YB. MLLD acknowledges \emph{Coordenação de Aperfeiçoamento de Pessoal de Nível Superior} - Brasil (CAPES) - Finance Code 001, \emph{Conselho Nacional de Desenvolvimento Científico e Tecnológico} (CNPq). KMD thanks the support of the CNPq. JHJ thanks to Brazilian institution CNPq for financial support through postdoctoral fellowship (project 150237/2017-0) and Chilean institution CONICYT, Programa de Astronom\'ia, Fondo ALMA-CONICYT 2017, C\'odigo de proyecto 31170038. ET was supported by ETAg grants IUT40-2, IUT26-2 and by EU through the ERDF CoE grant TK133 and MOBTP86. This research made use of the ``K-corrections calculator'' service available at http://kcor.sai.msu.ru/. \texttt{Python} wrapper of the \texttt{C} library by \citet{Blanton2007}, with \texttt{Python} wrapper on github\footnote{https://pypi.python.org/pypi/kcorrect\_python/2017.07.05}. Based on observations made with the JAST/T80 telescope at the Observatorio Astrof\'{\i}sico de  Javalambre (OAJ), in Teruel, owned, managed and operated by the Centro de Estudios de F\'{\i}sica del  Cosmos de Arag\'on. We acknowledge the OAJ Data Processing and Archiving Unit  (UPAD) for reducing and calibrating the OAJ data used in this work. Funding for the J-PLUS Project has been provided by the Governments of Spain and Arag\'on through the Fondo de Inversiones de Teruel; the Arag\'on Government through the Reseach Groups E96, E103, and E16\_17R; the Spanish Ministry of Science, Innovation and Universities (MCIU/AEI/FEDER, UE) with grants PGC2018-097585-B-C21 and PGC2018-097585-B-C22, the Spanish Ministry of Economy and Competitiveness (MINECO) under AYA2015-66211-C2-1-P, AYA2015-66211-C2-2, AYA2012-30789, and ICTS-2009-14; and European FEDER funding (FCDD10-4E-867, FCDD13-4E-2685). Funding for the SDSS and SDSS-II has been provided by the Alfred P. Sloan Foundation, the Participating Institutions, the National Science Foundation, the U.S. Department of Energy, the National Aeronautics and Space Administration, the Japanese Monbukagakusho, the Max Planck Society, and the Higher Education Funding Council for England. The SDSS Web Site is \url{www.sdss.org}. The SDSS is managed by the Astrophysical Research Consortium for the Participating Institutions. The Participating Institutions are the American Museum of Natural History, Astrophysical Institute Potsdam, University of Basel, University of Cambridge, Case Western Reserve University, University of Chicago, Drexel University, Fermilab, the Institute for Advanced Study, the Japan Participation Group, Johns Hopkins University, the Joint Institute for Nuclear Astrophysics, the Kavli Institute for Particle Astrophysics and Cosmology, the Korean Scientist Group, the Chinese Academy of Sciences (LAMOST), Los Alamos National Laboratory, the Max-Planck-Institute for Astronomy (MPIA), the Max-Planck-Institute for Astrophysics (MPA), New Mexico State University, the Ohio State University, University of Pittsburgh, University of Portsmouth, Princeton University, the United States Naval Observatory, and the University of Washington. This research is based on observations made with the Galaxy Evolution Explorer, obtained from the MAST data archive at the Space Telescope Science Institute, which is operated by the Association of Universities for Research in Astronomy, Inc., under NASA contract NAS 5$-$26555.

\end{acknowledgements}


\bibliographystyle{aa}
\bibliography{References_by_Nogueira-Cavalcante}

\end{document}